%% file: paper.tex
\documentclass[a4paper,conference]{IEEEtran}
\usepackage{cite}

\ifCLASSINFOpdf
   \usepackage[pdftex]{graphicx}
\else
\fi

\usepackage{amsmath}
\usepackage{amsmath}
\usepackage{amssymb}

\usepackage{array}
\usepackage{xspace}
\usepackage{multirow}
\usepackage{booktabs}
\usepackage{tabularx}

\usepackage{subcaption}
\usepackage[skip=4pt]{caption}

\usepackage{url}
\usepackage[pagebackref=true,breaklinks=true,letterpaper=true,colorlinks,bookmarks=false]{hyperref}

\usepackage{times}
\usepackage{epsfig}
\usepackage{xcolor}
\usepackage{animate}
\usepackage{nicefrac}

\newcommand*{\eg}{e.g.\@\xspace}
\newcommand*{\ie}{i.e.\@\xspace}
\newcommand{\etal}{{et al}.\@}

\begin{document}

\title{Motion Driven Video Generation}
\title{Rnn with auto-encoded states}
\title{One step at a time}
\title{Learning Directions for Single Image Video Synthesis}
\title{Learning Directions for Video Synthesis}
\title{Learning Directions}
\title{Learning to Take Directions One Step at a Time}



%
\author{
	\IEEEauthorblockN{
		Qiyang Hu\IEEEauthorrefmark{1},
		Adrian W\"alchli\IEEEauthorrefmark{1},
		Tiziano Portenier\IEEEauthorrefmark{1},
		Matthias Zwicker\IEEEauthorrefmark{2} and
		Paolo Favaro\IEEEauthorrefmark{1}
	}
	\IEEEauthorblockA{
		\IEEEauthorrefmark{1} University of Bern, Switzerland, \\
		\{qiyang.hu, adrian.waelchli, tiziano.portenier, paolo.favaro\}@inf.unibe.ch
	}%
	\IEEEauthorblockA{
		\IEEEauthorrefmark{2} University of Maryland, USA, 
		zwicker@cs.umd.edu
	}
}


\input{./figures/teaser.tex}

\begin{abstract}
We present a method to generate a video sequence given a single image. 
Because items in an image can be animated in arbitrarily many different ways, we introduce as control signal a sequence of motion strokes.
Such control signal can also be automatically transferred from other videos, \eg, via bounding box tracking.
Each motion stroke provides the \emph{direction} to the moving object in the input image and we aim to train a network to generate an animation following a sequence of such directions.
To address this task we design a novel recurrent architecture, which can be trained easily and effectively thanks to an explicit separation of past, future and current states. As we demonstrate in the experiments, our proposed architecture is capable 
of generating an arbitrary number of frames from a single image and a sequence of motion strokes.
Key components of our architecture are an autoencoding constraint to ensure consistency with the past and a generative adversarial scheme to ensure that images look realistic and are temporally smooth.
We demonstrate the effectiveness of our approach on the MNIST, KTH, Human3.6M, Push and Weizmann datasets. 
\end{abstract}


%
\IEEEpeerreviewmaketitle

\section{Introduction}

In this work we aim to train a neural network so that it can learn to ``take directions'' and thus make a moving object travel in a realistic way between two locations in an image.
In our framework directions are provided in the form of a sequence of \emph{motion strokes}, \ie, a rough 2D sketch of the path that a character in the input image should follow. We aim to generate an entire video with the desired animation by using only a single input image.  
%
As shown in Fig.~\ref{fig:teaser}, the input stroke can be extracted from a source video through bounding box tracking.

The key component in our proposed architecture is a novel recurrent neural network (RNN) augmented by an autoencoder that explicitly defines part of the state as the latent representation of a single video frame. 
The autoencoder shapes the latent representation space, which implicitly constrains the state transitions of the RNN.
We define the RNN input state as the encoding of: the initial frame, the current frame, the complete motion path and the next motion step. The RNN then predicts the next frame encoding. A fundamental benefit of this architecture is that the RNN can avoid degradation over time of the generated frames by relying on the encoding of the initial frame, can ensure temporal smoothness of the generated video by using the encoding of the current frame, and can produce a realistic animation by using the knowledge of the complete path. Moreover, because the next frame encoding does not need to accumulate the video and motion history (it is already part of the input), our proposed recurrent architecture is trained effectively with just a \textbf{single feedforward step} of the truncated back-propagation.
The predictor is trained by using videos where the bounding boxes of moving objects are available. 
Therefore, we can ensure coherence of the animation of a frame in a video by using a reconstruction loss between the output sequence of the RNN and the ground truth video. We also enforce image realism and temporal coherence by using adversarial training losses as well as perceptual losses. 
To the best of our knowledge, our work is the first to exploit successfully this RNN training strategy.

\subsection*{Contributions}
In summary, we introduce: 
\begin{enumerate}
	\item A novel \textbf{controllable video generation} from a single image and a sequence of motion strokes;
	\item A novel \textbf{recurrent network architecture and training scheme}, which allows single-step truncated back-propagation through time (see Section~\ref{sec:training});
	\item Demonstrate controllability of the trained model on the MNIST, KTH, Human3.6M, Push and Weizmann datasets with higher performance than competing video-generation methods.
\end{enumerate}

\section{Related Work}\label{sec:related-work}
Early works on video synthesis focused on generating continuously varying textures, the so-called \emph{video textures}, from single or multiple still frames~\cite{schodl2000video, yin2011shape}.
Recently, deep generative models such as Generative Adversarial Networks (GAN) or variational autoencoders (VAE) have been successfully used to generate realistic images or videos from latent codes~\cite{goodfellow2014generative, kingma2013auto, saito2017temporal, vondrick2016generating}.
Furthermore, even higher quality results have been obtained by using models conditioned on additional input, \eg, the class label, the content image or the semantic label map~\cite{mirza2014conditional, Pan_2019_CVPR, van2016conditional}.
Recurrent neural networks~\cite{hochreiter1997long} are a natural choice to learn from time-dependent signals such as text, video or audio.
For example, Tulyakov \etal~\cite{tulyakov2018mocogan} decompose motion- and content components directly in the latent space. They use a recurrent GAN architecture and sample the motion vector at each time step.
Byeon \etal~\cite{byeon2018contextvp} instead use a multi-dimensional Long Short-Term Memory (LSTM) that aggregates contextual information in a video for each pixel over space and time.
In this work, we also train an RNN to generate a video conditioned on a still frame and a sequence of motion strokes. 
RNNs are known to be difficult to train compared to feedforward neural networks~\cite{pascanu2013difficulty}.
However, thanks to our proposed design, training can be done successfully even with a single time step truncated back-propagation.
Moreover, to the best of our knowledge this is the first work that uses 2D strokes as a motion representation for animation in a generative setting.
In the following paragraphs, we describe work related to video prediction and motion editing.

\paragraph{Video Prediction from Multiple Frames}
It is well-known that using the mean-squared error as the reconstruction loss leads to blurry frame predictions.
To obtain sharper image predictions, Mathieu \etal~\cite{mathieu2015deep} combine the $\ell^2$-loss with a GAN objective. 
They do not provide input noise to the generator, however, and thus their predictions are fully deterministic. 
Denton \etal~\cite{denton2018stochastic, denton2017_disentagled} learn the prior distribution of the latent space at each time step of their LSTM given the previous frame, and sample from the learned prior to predict the next frame.
In contrast, we generate all future frames from a single frame and an encoding of the motion strokes.
Instead of multiple starting frames, Wang \etal~\cite{Wang_2019_ICCV} use start- and end-frames as targets and generate a smooth motion between them with, however, limited control of the motion path.  

\paragraph{Video Prediction from a Single Frame}
Predicting the future from a single still image is highly ambiguous.
Most prior work uses a variational approach in order to constrain the future outcomes in the training phase and at the same time have the possibility to sample from the latent space at test-time~\cite{babaeizadeh2018stochastic, hao2018controllable, he2018probabilistic, li2018flow}. For example, Xiong \etal~\cite{xiong2018learning} use a GAN and a Gram matrix for motion modelling. 

There are three works that closely relate to ours. The first two,
Li \etal~\cite{li2018flow} and the concurrent work of Endo \etal~\cite{Endo19}, predict a video from a single image using a variational autoencoder to sample optical flows conditioned on the input frame. 
A separately trained network then synthesizes the full-frames from the optical flow maps. 
In contrast, we provide only a single motion stroke sequence, which corresponds to the average optical flow of the object of interest. 
While optical flow is a precise description of how objects move from frame to frame, it is also very specific to the object instance. 
We believe that a motion stroke sequence is a much more transferrable motion representation across videos.
The third related work, Hao \etal~\cite{hao2018controllable}, synthesizes a video clip from a single image and a matrix of trajectories. 
It generates a video of the whole scene, whereas we focus on animating the object in the image. 
Moreover, due to its recurrent design, our proposed architecture is able to output a variable length video. 


\input{./figures/architecture.tex}

\paragraph{Human Motion Synthesis}
Recent work in motion synthesis has demonstrated the effectiveness of convolutional neural networks for accurate human pose estimation in real time~\cite{cao2017realtime, newell2016stacked, wei2016convolutional}.
The pose extraction from real images can be used further to synthesize images of people in novel poses. 
Balakrishnan \etal~\cite{balakrishnan2018synthesizing} achieve this by segmenting the persons body parts and background, by transforming the parts to the new pose and by fusing the result with the background.
Chan \etal~\cite{chan2019everybody} transfer the pose of a person in a source video to a target subject in another video. They enforce the target motion with skeletal poses as input.
Both of these works can render high-quality video with realistic motions because of the strong supervision through pose. 
In our case with a single motion stroke sequence, we are given a very sparse description of motion and lack the exact location of all limbs. 
Hence, in our case there is more ambiguity and a bigger challenge to generate realistic renderings. On the other hand, we do not require as much detailed supervision (\eg, sketches for training can be extracted automatically via bounding box tracking).

\paragraph{Sketch-based Animation}
To this date, character animation remains a challenging and labor-intensive task.
Early works for automated animation from sketches focused on cartoon figures, which, despite their simplistic appearance, have a similar complexity in terms of motion compared to real images. 
Davis \etal~\cite{davis2003sketching} take a sequence of 2D pose sketches and reconstruct the most-likely 3D poses which are applied to a 3D character model for animation.
Thorne \etal~\cite{thorne2004motion} require only a sketch of the character and a continuous stroke for the motion. 
Chen \etal~\cite{chen2005character} reconstruct the 3D wireframe of the character from a sequence of sketches and correspondences. This allows them to add realistic lighting, textures and shading on top of the animated character.
Our system is fully end-to-end and does not require to explicitly model the 3D or rendering pipeline, and apart from the motion stroke sequence no other annotation is needed.


\section{Video from a Motion Stroke Sequence}\label{sec:approach}
We formulate our problem as follows: Given a single image $I_0\in \mathbb{R}^{m\times n \times c}$, with $m\times n$ pixels and $c$ color channels, containing a movable object, and a  motion stroke $S\in [0,1]^{m\times n}$, we aim to synthesize $T$ future images $I_1, \dots ,I_T$, \ie, a video, where the object moves realistically along the provided stroke.
We assume that the stroke starts roughly at the center of 
the object's bounding box.
We focus on human motion 
because it provides a challenging playground with complex non-rigid deformations, although our model does not make any assumptions specific to humans and indeed we demonstrate it also on videos of a robotic arm in the Experiments section.



\subsection{A One-Step Recurrent Architecture}


To enable the synthesis of variable length videos, we address the problem with a recurrent neural network that continually outputs video frames.
Our network is composed of three main stages that are jointly trained in an end-to-end fashion: The encoding stage, the prediction stage, and the decoding stage (see Fig.~\ref{fig:architecture}).
We use two encoders to extract texture information in the input frame and motion information from the stroke sequence.
These are concatenated and fed to our predictor, which outputs the encoding of the next frame.
At test time this predictor is applied recursively by feeding the output $\hat h_{t+1}$ back as input $\hat h_{t}$.
In addition, the encoding of the initial frame is always given as input to retain a reference to the beginning of the sequence.
At the decoding stage the generator network outputs one RGB frame at a time.
In the following paragraphs we detail each of these building blocks and point out differences at training and test time.
For clarity, we omit the notation for network parameters from our formulations.

\paragraph{Encoding Stage}
At the beginning we are given the initial frame $I_0$ and motion stroke sequence $S$.
First, we encode the image $I_0$ with $E_1$ to a feature $h_0$.
For the motion input, we extract every consecutive one-step stroke $S_t$ between keypoint $t$ and $t+1$ from $S$ and concatenate the two to obtain the \emph{instant motion} encoding 
\begin{align}
x_t = E_2(S, S_t).
\end{align}
In the supplementary material, we show experimentally that both $S$ and $S_t$ are needed.
Intuitively, since $S$ provides the full motion, the system is given a conditioning of the future at each time step, while the current motion step $S_t$ explicitly tells the system the current time and velocity of motion.
Only providing $S_t$ would leave too much ambiguity and would not allow the predictor to relate the current frame encoding to the encoding of the initial frame.

\paragraph{Prediction and Decoding}
At any time step $t>0$, we input the encodings $h_0$ and $x_t$ to our recurrent predictor $P$
\begin{equation}
	\hat{h}_{t+1} = P(h_0, \hat h_t, x_t)
\end{equation}
and obtain the next state $\hat h_{t+1}$. When $t=0$ we use $\hat{h}_{1} = P(h_0, h_0, x_0)$.
Note that, at any time, the predictor has not only access to the instant motion, but also to the entire stroke sequence $S$, which was given as input.
Finally, each output $\hat{h}_{t+1}$ of the predictor is individually decoded by $G$ to produce RGB frames $\hat{I}_{t+1}$ that are then concatenated to form the final video. 
	

\subsection{Training}\label{sec:training}
We train all parts of the network together by minimizing a combination of reconstruction, adversarial and perceptual losses.
Firstly, we use an $\ell^2$-loss between ground truth $I_{t+1}$ and synthesized frame
$\hat{I}_{t+1} = G(\hat{h}_{t+1})$, and an autoencoding constraint between $E_1$ and $G$, \ie, 
\begin{align}\label{eq:rec-loss1}
	\mathcal{L}^1_\text{rec} &= \sum_{t=0}^{T} \, \lvert \hat{I}_{t+1} - I_{t+1}\rvert_2,
	\\ 
	\mathcal{L}^2_\text{rec} &= \sum_{t=1}^{T} \, \lvert \hat{I}_{t} - G(E_1(I_{t}))\rvert_2.
\end{align}
This choice of loss function encourages the output to match the ground-truth \textbf{in expectation}, which allows flexibility in the input-output mapping.
Furthermore, to avoid blurred outputs from the pixel-wise loss, we use two discriminators: 
One that distinguishes between predicted fake frames $\hat{I}_{t+1}$ and real frames $I$ from the distribution of real images $p_\text{data}$, \ie,
\begin{equation}\label{eq:gan-loss-1}
	\mathcal{L}_\text{GAN}^1 = 
	\mathbb{E}_{I} 
	\left[\log D_1(I)\right] + \mathbb{E}_{I_0,S,S_t} 
	\left[\log (1 - D_1(\hat{I}_{t+1}))\right], 
\end{equation}
and another one that discriminates generated pairs $\hat{q}_t = (\hat{I}_t, \hat{I}_{t+1})$ from real pairs $q_t = (I_t, I_{t+1})$ of consecutive frames, \ie 
\begin{equation}\label{eq:gan-loss-2}
	\mathcal{L}_\text{GAN}^2 = 
	\mathbb{E}_{q_t} 
	\left[ 
		\log D_2(q_t)	\right] + \mathbb{E}_{I_0,S,S_t} 
	\left[\log (1 - D_2(\hat{q}_t))
	\right].
\end{equation}
The latter GAN loss leads to a temporal smoothing and is inspired by the corresponding term used in Chan \etal~\cite{chan2019everybody}.
Furthermore, the single-frame discriminator in eq.~\eqref{eq:gan-loss-1} is conditioned on the instant motion encoding $x_t$, which we omitted from the notation for better readability.
We also use the perceptual loss~\cite{johnson2016perceptual}
\begin{equation}
\mathcal{L}_\text{VGG} = \sum_{t=0}^{T} \, \lvert \Phi(\hat{I}_{t+1}) - \Phi(I_{t+1})\rvert_2
\end{equation}
between output video and ground truth video, where
$\Phi$ are the features extracted from a pre-trained VGG network~\cite{simonyan2014very}.
Finally, we formulate the overall objective as
\begin{align}\label{eq:gan-objective}
	\min_{\theta_1} \max_{\theta_2}
	&
	~\mathcal{L}_\text{GAN}^1\! +\! 
	\lambda_0 \mathcal{L}_\text{GAN}^2\! +\! 
	\lambda_1 \mathcal{L}_\text{VGG}\! +\! 
	\lambda_2 \mathcal{L}^1_\text{rec}\! +\!
	\lambda_3 \mathcal{L}^2_\text{rec}\!
\end{align}
where $\theta_1 = \left\{\theta_{E_1}, \theta_{E_2}, \theta_P, \theta_G \right\}$ denote the network weights for the encoders and the generator, $\theta_2 = \left\{\theta_{D_1}, \theta_{D_2}\right\}$ denote the network weights for the discriminators, and $\lambda_0,\dots,\lambda_3>0$ are tuning parameters.
We optimize the above objective by alternating stochastic gradient descent on $\theta_1$ and $\theta_2$.

\subsection{Single Time Step Training}
Since the predicted frame at time $t+1$ depends on the input at time $t$, we train the network sequentially by using the previous output as input (see red arrow in Fig.~\ref{fig:architecture}).
The encoding $h_t$ can be regarded as the state of the system at time $t$ and $x_t$ is the input.
In general, for recurrent neural networks the state $h_t$ can only be computed by knowing the state $h_{t-1}$~\cite{hochreiter1997long}.
This is achieved by \emph{unrolling} the RNN and applying back-propagation through time (BPTT).
In our unique setting, BPTT is not necessary as in each step all information required to predict the next state is included in the input and there is no need to accumulate the sequence history in the predicted state. 
It allows us to train our predictor as a feed-forward network (no weight sharing over time) despite the fact that $P$ has a recurrent nature. 
As an additional benefit, initializing the state at time $t=0$ is straightforward, unlike in regular RNNs where typically the state is initialized randomly, with a constant, or with a learned parameter.

\input{./figures/teacher_forcing.tex}

\paragraph{Teacher Forcing Training Fails}

An alternative training method for our architecture, called \emph{teacher forcing}, is to directly feed the encoding $h_t = E_1(I_t)$ to the predictor $P$, rather than using the feedback from the previous output $\hat h_t$ of $P$. However, this technique fails to train the network. Fig.~\ref{fig:results-kth-fail} shows two (failure) examples of generated videos after using this training technique. The network ignores the motion input (the same video is generated for two different motion stroke sequences) and also it ``forgets'' the initial input image (the future frames become darker and darker).

\subsection{Runtime}
At test time our system is given only a single image $I_0$ and the stroke sequence $S$. 
The encoders $E_1$ and $E_2$ are used at time $t=0$ to obtain encodings $h_0$ and $x_0$ on which the predictor $P$ is recursively applied for a number of time steps.
At each time step, the predicted next encoding $\hat h_{t+1}$ is fed back as input for the next step that produces $\hat h_{t+2}$ and so on.
As in the training phase, $h_0$ is given as input at each step to provide a reference point to the beginning of the sequence.

\subsection{Implementation}
We use convolutional layers for encoding and transposed convolutions for decoding.
The predictor $P$ is composed of dense-blocks~\cite{huang2017densely}.
Spectral normalization~\cite{miyato2018spectral} is only employed in the discriminator and no other normalization technique is applied.
The spatial dimensions of the input and output are $128 \times 128$ pixels.
We use the Adam optimizer~\cite{kingma2014adam} with learning rate $2 \cdot 10^{-4}$ and $\beta_1 = 0.5$ for both the generator and discriminator. 
The hyperparameters in eq.~\eqref{eq:gan-objective} for all the experiments are: $\lambda_0 = 1$, $\lambda_1 = 10$, $\lambda_2,\lambda_3 = 20$. 
We find that our model is not very sensitive to the choice of hyperparameters.
Further details about the architecture of our network are presented in the supplementary material.
\footnote{Code available at:~\url{https://github.com/HuQyang/learning-direction}}

\input{./figures/mnist_3.tex}

\section{Experiments}\label{sec:experiments}

We evaluate our models using several datasets: the MNIST~\cite{lecun1998gradient} handwritten digits, the robot Push dataset~\cite{pushdataset}, the KTH human actions~\cite{schuldt2004recognizing}, the Weizmann institute dataset~\cite{ActionsAsSpaceTimeShapes_pami07} and Human3.6M~\cite{ionescu2014human3} (see supplementary material). 
We show qualitative results on all datasets and quantitative evaluations on both robot push and KTH and ablation studies on KTH. 
The results on MNIST and Push demonstrate that our method can potentially generate videos with very diverse motion stroke sequences as input. 
Our method not only translates the stroke sequence to an object trajectory, but also accurately maps the velocity encoded in the stroke sequence to the output video.
Unlike the rigid robot arm or MNIST digits, the KTH, Weizmann and Human3.6M datasets consist of non-rigid, realistic human actions and are therefore more challenging. 
We mainly show three typical human motions with different speeds: walking, running and jogging.
The results on these three datasets show that our method is efficient in synthesizing realistic videos involving complex human motions.

\subsection{Stroke Sequence Generation}
For MNIST, the trajectory is randomly generated online during training and testing, and for Push we manually select 14 patches of size $64 \times 64$ pixels from different videos and we search in the first frame in each video the patch most similar to the initial one by computing the perceptual loss~\cite{johnson2016perceptual} between the reference patch and the searched patch. 
We then use the initial patch to search the next frame. 
A single stroke point is defined as the center of a patch.
For the human actions datasets, as an alternative to annotating the training set manually, we compute the bounding box with the YOLO object detector~\cite{redmon2016you}, track its center in each frame and use it as the stroke point. 

In all cases, we encode the time instant in each stroke with a pixel's grayscale intensity (black indicates the beginning and light grey indicates the end). At test time one stroke is extracted the same way as in training, but applied to an image taken from a different video.

\subsection{MNIST}
The MNIST dataset~\cite{lecun1998gradient} consists of $60$k handwritten digits for the training and $10$k for the test set. 
In order to test our system on arbitrary trajectories, we create a synthetic dataset of moving MNIST digits. 
We take the $28 \times 28$ digit images and move them within a window of $112 \times 112$ pixels for 16 frames. 
The first frame is given as the input and the system predicts the following 15 frames. 
Our method can generate video sequences from one single image coherent with the input stroke sequence. 
We demonstrate this in two ways in Fig.~\ref{fig:results-mnist}: The first two rows show examples where the stroke sequence is changed and the digit is kept the same; in the remaining rows the same stroke sequence is applied to different digits.

\subsection{Robot Push dataset}
The Push dataset~\cite{pushdataset} consists of $\sim$59k videos of robot interactions involving pushing motions, including one training set and two test sets of previously seen and unseen objects.  
We show our results and qualitative comparison with Hao \etal~\cite{hao2018controllable} in Fig.~\ref{fig:results-robot-qual}. 
We only show the last (8-th) predicted frame to compare with them. One can see that our method produces sharper and realistic-looking images.
In Fig.~\ref{fig:results-robot-quan} we evaluate our results quantitatively over time using the common similarity metrics PSNR and SSIM \cite{ssim}, and find that we outperform Hao \etal~\cite{hao2018controllable}. 
In the supplemental material we show the first 8 predicted frames as done in \cite{hao2018controllable}.

\input{./figures/push.tex}
\input{./figures/push_psnr_ssim.tex}

\subsection{KTH}
The KTH Action dataset~\cite{schuldt2004recognizing} contains grayscale videos of 25 persons performing various actions from which we choose the subsets \emph{walking}, \emph{running} and \emph{jogging} since the others do not contain large enough global motions. 
We train on all the selected data except persons 21-25 which we reserve for testing, yielding $\sim$98k and $\sim$19k frames for training and testing respectively.
Since there are sequences where the person is walking out of the frame, we employ the YOLO object detector~\cite{redmon2016you} to exclude frames where the confidence level of detecting a person is less than~$\nicefrac{1}{2}$.

\paragraph{Qualitative Results} 
Examples of synthesized sequences are shown in Fig.~\ref{fig:results-kth}.
We randomly pick a subsequence of 17 frames from the dataset of videos and treat it as one video clip. 
We use the first frame as input and compute the center of the bounding box of each frame to get the stroke sequence. 
The system predicts the following 16 frames. 
In order to demonstrate that the output of our network is controllable through the motion stroke sequence, we test it with varying input stroke sequences $S$ while keeping the initial frame $I_0$ fixed. 
Examples of these tests are shown in Fig.~\ref{fig:results-kth} for the KTH dataset~\cite{schuldt2004recognizing}. 
The stroke sequence in each row is taken from a different video in the test set and applied to the image in column one. We can see that given the same input image, different stroke sequences yield a different video. 
It turns out that the intervals between two stroke points encode the information of the motion. 
Denser stroke sequences are translated into walking motion and stroke sequences with large intervals correspond to running motion. 
Between these two stroke sequence types is the jogging motion. We also observe that the generated frames follow the position of the stroke points. 
Furthermore, if the pose of a person in the input image resembles a running posture, but the stroke sequence describes a walking motion, the system generates a realistic and smooth change from running to walking, instead of jumping directly to a walking pose. 
The generated video clips are shown in the supplemental material.
\input{./figures/weizmann.tex}
\paragraph{Long Sequences}
Since our system is generating images recurrently frame by frame, we are not bound by the length of videos seen during training.
In Fig.~\ref{fig:results-kth-long-seq} we test with longer stroke sequences of 24 keypoints, yielding 24 synthesized frames.
Even though we trained with videos to predict 16 frames, this experiment shows that at test time we can generate longer sequences and maintain the sharpness.
We speculate that if the system is trained with longer sequences, then it could generate even more high quality frames.

\input{./figures/kth_2.tex}
\input{./figures/long_sequences.tex}
\input{./figures/comparison_li_endo.tex}

\paragraph{Qualitative Comparison}
There are also many video prediction methods requiring multiple frames as input. 
We take the recent work from Denton \etal~\cite{denton2018stochastic} and compare with them. 
Since they require 10 frames as input, we take their 10th frame as our input. The figures are shown in the supplemental material.
We observe that our result with only one image is comparable to Denton \etal~\cite{denton2018stochastic} with 10 frames as input.
We also compare to Li \etal~\cite{li2018flow} who, as in our method, require only one image as input, but instead of a motion stroke sequence they provide a noise vector as second input, which makes their method not directly controllable.
They predict a fixed number of 16 frames.
Fig.~\ref{fig:comparison-all} shows that we generate images with a quality similar to that of Li \etal~\cite{li2018flow}.
In the same figure we also compare with the concurrent work of Endo \etal~\cite{Endo19}.


\paragraph{Quantitative Analysis} 
We perform a motion statistics evaluation on the generated sequences, and use the Learned Perceptual Image Patch Similarity (LPIPS)~\cite{zhang2018unreasonable} as Li \etal~\cite{li2018flow} do, to test the realism of the generated images.
Since video prediction in our and the compared work is non-deterministic, evaluation using a pixel-wise metric is not meaningful.
Instead, we compute motion and object detection statistics on generated as well as ground-truth KTH sequences independently.
We extract the pose joints in consecutive images using the convolutional pose machine~\cite{wei2016convolutional} and measure the mean and standard deviation of the Euclidean distance between corresponding joints.
In Table~\ref{tbl:quantitative-motion}, lower numbers represent a smoother trajectory of detected pose joints, meaning that the pose detection is benefitting from a better image quality.
The pose is only evaluated on the subset of frames where a person is detected, and the fraction of these is listed in the last column.
The detection rate on the ground truth is 100\% since we train and test only on frames where a person is detected.
Table~\ref{tbl:quantitative-motion} shows that we outperform in all motion categories except \emph{running} and have the highest object detection rate.
Furthermore, we evaluate the LPIPS~\cite{zhang2018unreasonable} perceptual metric for 16 predicted frames. We re-train the publicly available code of Li \etal~\cite{li2018flow} and Denton \etal~\cite{li2018flow} on our data.
We achieve 0.09, which is the best score compared to \cite{li2018flow} (0.14) and \cite{denton2018stochastic} (0.11).


	

\input{./figures/pose_smoothness.tex}

\subsection{Weizmann}
We further train our system on the Weizmann institute dataset \cite{ActionsAsSpaceTimeShapes_pami07} with motion types \emph{walking, running, jumping, galloping, bending} and \emph{skipping}, and $64 \times 64$ pixels images.
Qualitative results are shown in Fig.~\ref{fig:weizmann1}. 
As in the other experiments, we randomly pick test stroke sequences from different videos to animate a person in a selected image.

%

\section{Conclusions}\label{sec:conclusion}
In this paper, we propose a novel method to synthesize a video clip from one single image and a motion stroke sequence. 
Our method is based on a recurrent architecture and is capable of generating videos by predicting the next frame given the previous one. 
The way we formulate the problem allows us to explicitly describe the state of the system and apply a novel training scheme for RNNs that eliminates the need for the costly back-propagation through time.
Hence, our system can be trained efficiently and still generate videos of arbitrary length. 
We demonstrate this on several real datasets with rigid as well as non-rigid motion and find that it can animate images realistically.
The model is able to generalize on in-distribution motion trajectories while the rendering quality degrades for out-of-distribution motion trajectories. 
We believe that this limitation could be addressed by collecting and training on more data. 
Moreover, to increase the range of its applications, the model could be extended to the generation of a moving background. In this regard, our work is a first fundamental building block towards this more general setting, which we leave for future work.



\bibliographystyle{IEEEtran}
\bibliography{egbib}

\end{document}

%% file: figures/teaser.tex
\twocolumn[{%
	\renewcommand\twocolumn[1][]{#1}%
	\maketitle
	\vspace{-.5cm}
	\newcommand{\rowheight}{1.2cm}
	\newcommand{\sepspace}{0.9cm}
	\begin{minipage}{\linewidth}
		\setlength\tabcolsep{1mm}
		\def\arraystretch{0.8}
		\begin{tabular}{llll}
			\multirow{3}{*}{\hspace{0.5cm}\includegraphics[height=3.2cm]{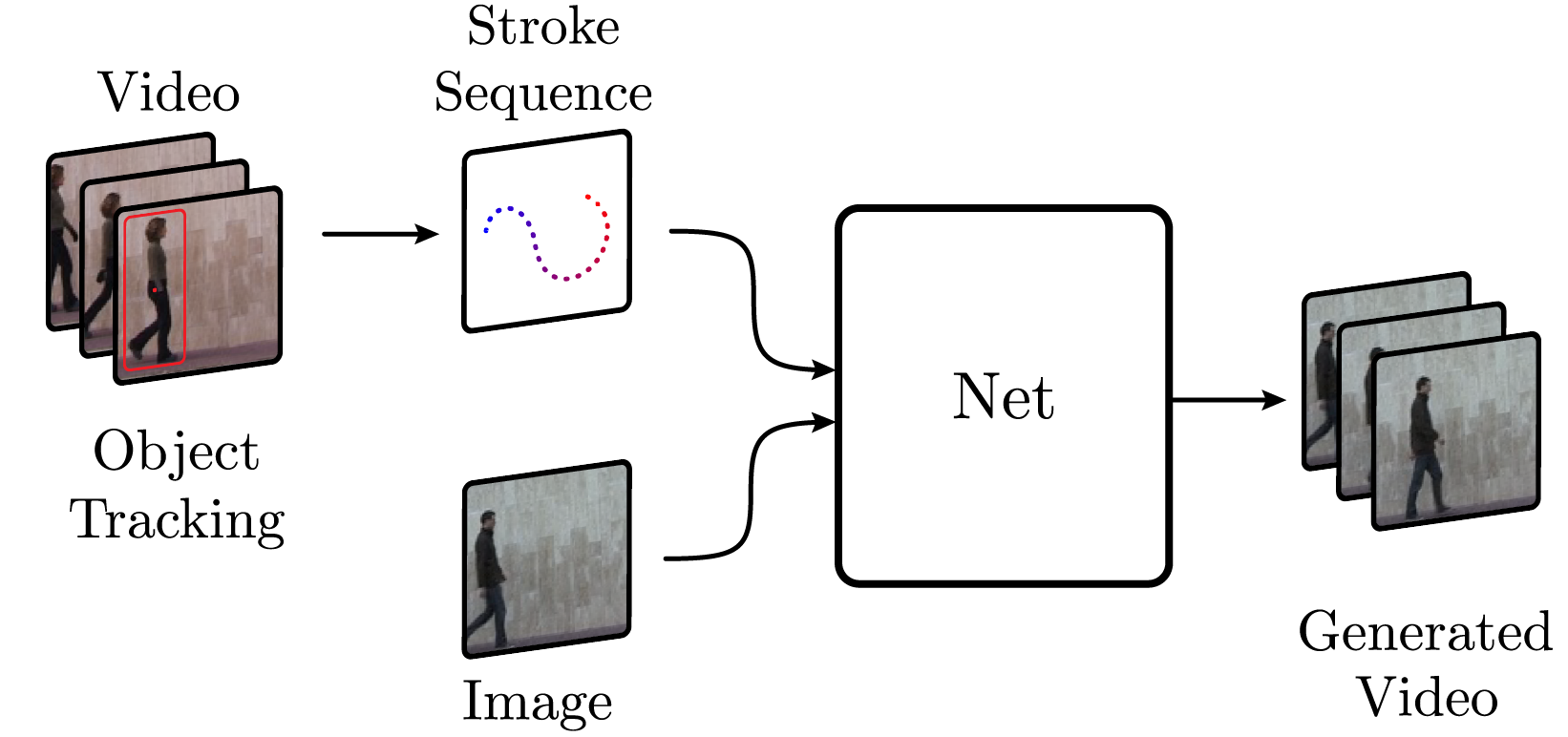}}& 
			\hspace{\sepspace}{\small Stroke} & {\small Input}& {\small Generated}
			\\
			\multirow{3}{*}{} &
			\hspace{\sepspace}{\small sequence} & {\small image} &  {\small frames} 
			\\
			& \hspace{\sepspace}
			\includegraphics[height=\rowheight]{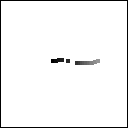} &
			\includegraphics[height=\rowheight]{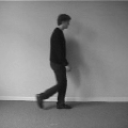} &
			\includegraphics[height=\rowheight]{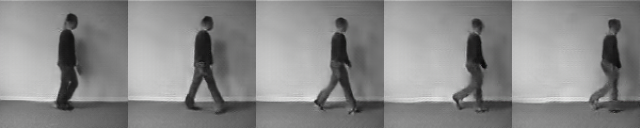} 
			\\
			& \hspace{\sepspace}
			\includegraphics[height=\rowheight]{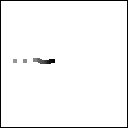} &
			\includegraphics[height=\rowheight]{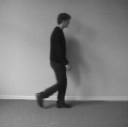} &
			\includegraphics[height=\rowheight]{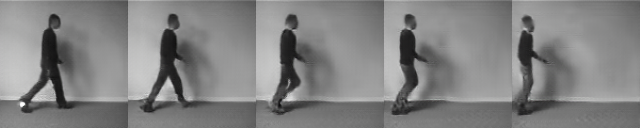} 
			\\
			\multicolumn{1}{c}{(a)} & \multicolumn{3}{c}{(b)}
		\end{tabular}
	\end{minipage}~%
	\captionof{figure}{\label{fig:teaser}%
		(a) Our network learns to take directions in the form of motion strokes extracted from a video through object tracking and generate videos from a single image. 
		(b) Different animations with the same initial image: (top row) forward motion and (bottom row) backward motion.
		Videos are shown in the supplementary material.
	}\vspace{.5cm}
}]

%% file: figures/architecture.tex
\begin{figure*}[t]
	\centering
	\includegraphics[width=0.93\linewidth]{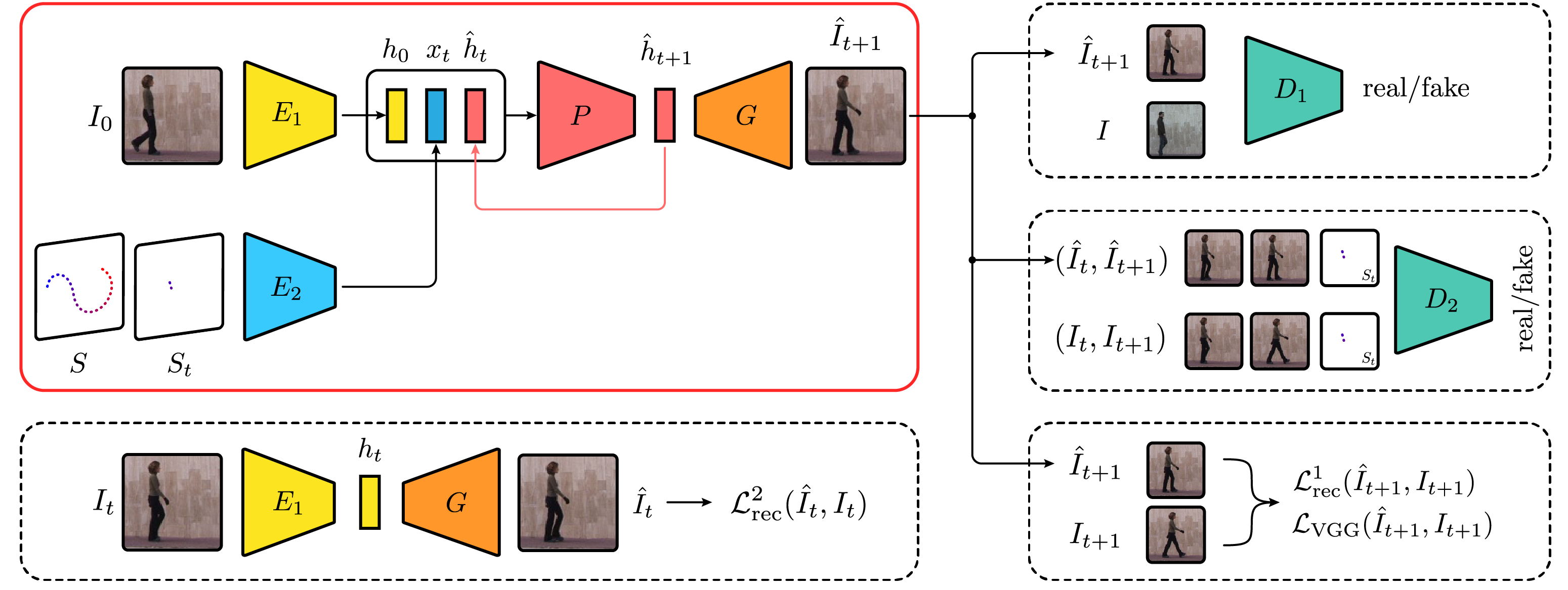}
	\caption{
		\label{fig:architecture}
		 \textbf{Network architecture.} 
		 The main network is inside the red box and loss terms are enclosed by dashed boxes. 
		 Our predictor $P$ recurrently predicts the state $\hat h_{t+1}$ of a new frame given the previous one, $\hat h_{t}$, the initial state $h_{0}$ and an encoding $x_{t}$ of the motion stroke sequence (current motion stroke, $S_t$, and complete motion, $S$). 
		 Using encoders $E_1$ and $E_2$, the predictor leverages the learned encoding $h_0$ of the input image $I_0$, the motion stroke sequence $S$ and the current motion step $S_t$, as additional inputs. 
		 Given these inputs, the predictor $P$ can preserve the appearance of the original image $I_0$ and make the animated object follow the input stroke sequence.
		 The predicted encoding is decoded into a temporally consistent image frame $\hat{I}_{t+1}$ using the generator $G$, which is trained both with reconstruction constraints $\mathcal{L}^1_{\text{rec}}$ and $\mathcal{L}_{\text{VGG}}$, and in an adversarial manner by competing against discriminators $D_1$ (for image realism) and $D_2$ (for temporal smoothness).
		 Finally, the encoding space of $E_1$ and generation ability of $G$ are further constrained and trained through the autoencoding loss $\mathcal{L}^2_{\text{rec}}$.
	}
	\vspace{-2mm}
\end{figure*}

%% file: figures/teacher_forcing.tex
\begin{figure}
   	\centering
	\setlength\tabcolsep{2pt}
	\newcommand{\height}{2.8cm}
	\def\arraystretch{0.8}
	\begin{tabular}{lll}
		{\small Stroke} & {\small Input }& {\small Generated }\\
		{\small sequence} & {\small image } & {\small frames }\\
		\includegraphics[height=\height]{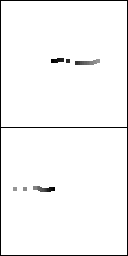} &
		\includegraphics[height=\height]{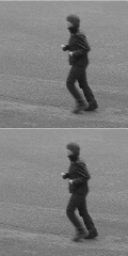} &
		\includegraphics[height=\height]{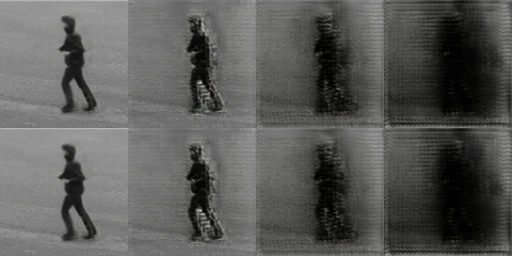} \\
	\end{tabular}
	    \caption{
	   		\label{fig:results-kth-fail} 
	   		Failure cases when training our architecture with teacher forcing. 
	   		Shown are the first four generated frames.
	    }
\end{figure}

%% file: figures/mnist_3.tex
\begin{figure*}
	\centering
	\small
	\def\arraystretch{0.1}	
	\newcommand{\height}{4cm}
	\newcommand{\vheight}{0.246\height}
	\newcommand{\vsep}{0.1mm}

		\setlength\tabcolsep{1pt}
		\begin{tabular}{lll}
			{\small Strokes} &  {\small Input} & \small{Generated frames} \\
			\includegraphics[height=\height]{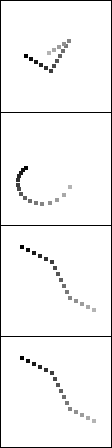}  \hspace{3mm} &
			\includegraphics[height=\height]{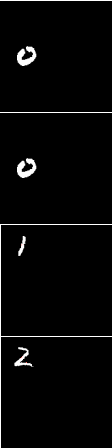}  \hspace{3mm} &
			\includegraphics[height=\height]{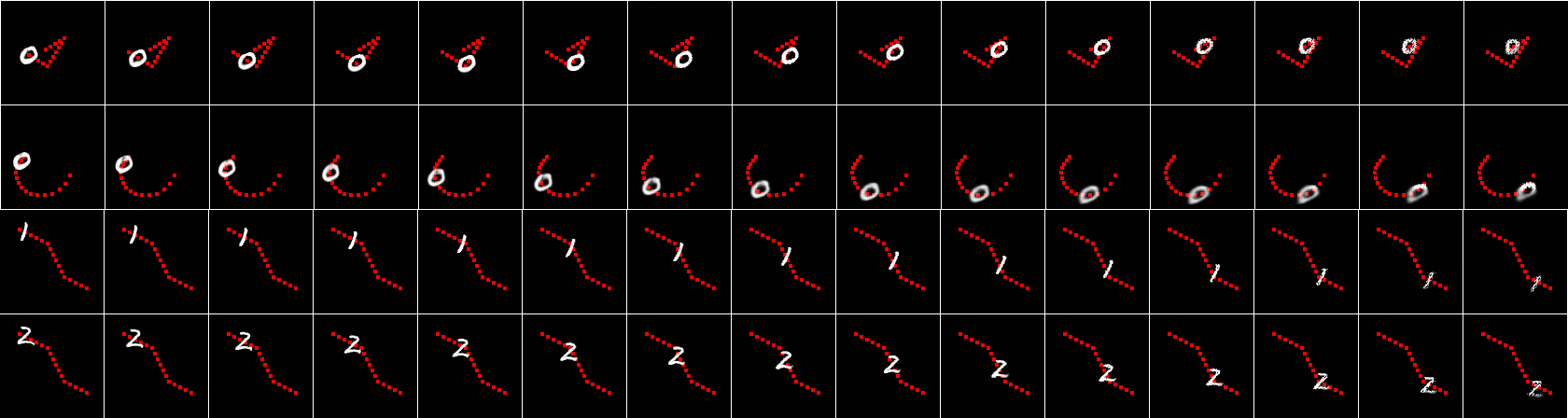} 
		\end{tabular}%
		

    \caption{
	    \label{fig:results-mnist}
	    Results on the MNIST dataset: Same motion stroke applied to different digits and of different motions applied to the same input frame. 
	    First column: The input stroke sequence, where the intensity of the dots goes from black to light grey with increasing time. Second column: The input image. 
	    The remaining columns show the generated video frames. 
	}
    \vspace{-1mm}
\end{figure*}

%% file: figures/push.tex
\begin{figure}[t]
	\centering
	
	\begin{subfigure}[b]{\linewidth}
	\begin{tabular}{p{0.18\linewidth}p{0.22\linewidth}p{0.20\linewidth}p{0.20\linewidth}}
		{\small Input} & {\small Hao~\etal~\cite{hao2018controllable}} & {\small Ours}  & {\small Ground truth}\\
	\end{tabular}
		\includegraphics[width=\linewidth]{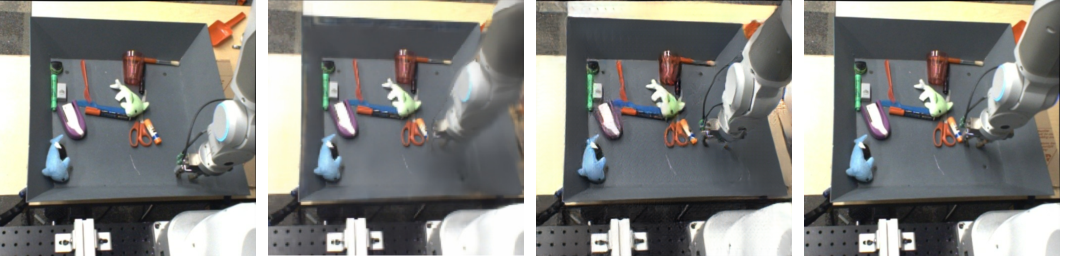}
		\includegraphics[width=\linewidth]{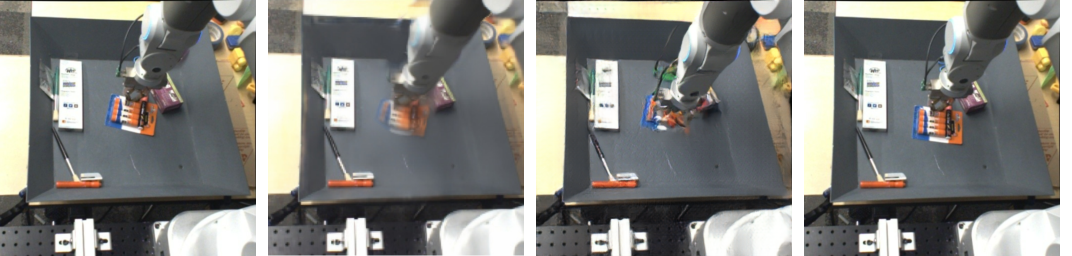}
		\includegraphics[width=\linewidth]{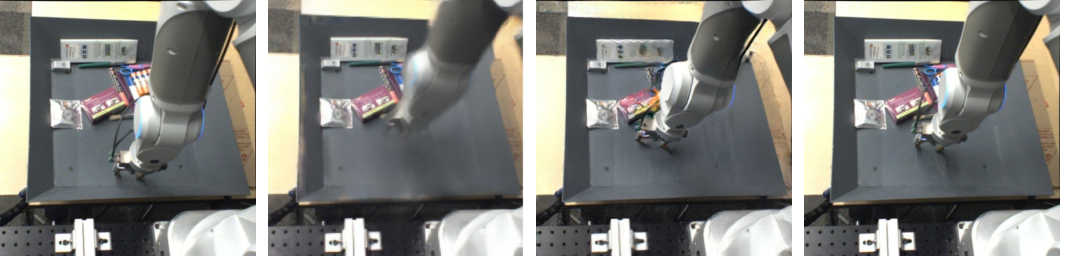}
	
	\end{subfigure}
	\caption{
		\label{fig:results-robot-qual}
		Qualitative results on the Push~\cite{pushdataset} test set with novel scenes and motions in comparison with \cite{hao2018controllable}.
	} 
\end{figure}

%% file: figures/push_psnr_ssim.tex
\begin{figure}[t]
	\centering
	
	\begin{subfigure}[b]{0.5\linewidth}
		\includegraphics[width=\linewidth]{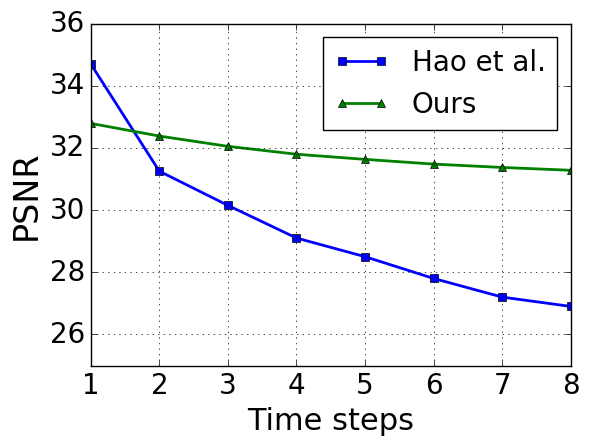}
	\end{subfigure}%
	\begin{subfigure}[b]{0.5\linewidth}
			\includegraphics[width=\linewidth]{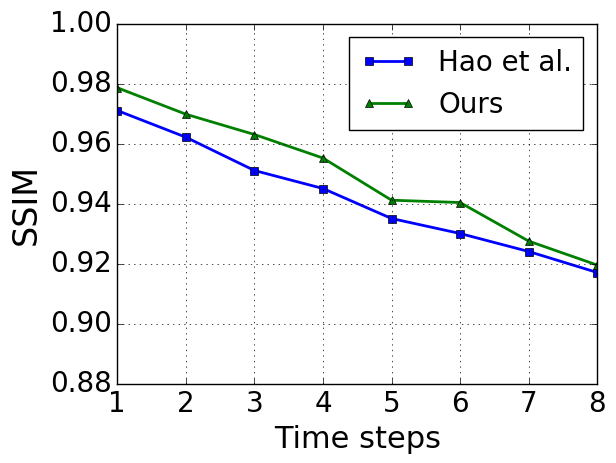}
		\end{subfigure}
	\caption{
		\label{fig:results-robot-quan}
		PSNR \& SSIM over time on the Push dataset \cite{pushdataset} in comparison with Hao \etal~\cite{hao2018controllable}.
	} 
\end{figure}

%% file: figures/weizmann.tex
\begin{figure}[t]
	\centering
	\includegraphics[width=\linewidth]{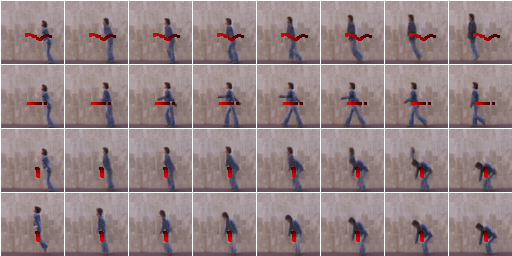}
	\caption{
		\label{fig:weizmann1}
		Four animations on the Weizmann dataset~\cite{ActionsAsSpaceTimeShapes_pami07}. 
		Every other column is skipped. 
		Top to bottom: jumping, walking, bending and landing.
		The stroke is shown in red. 
		Rows 1-3 have the same input image and different motion stroke sequences.
		Rows 3 and 4 have different input images, but the same motion stroke sequence.
	}
\end{figure}

%% file: figures/kth_2.tex
\begin{figure*}
	\begin{subfigure}[b]{\textwidth}
		\def\arraystretch{0.8}
		\setlength\tabcolsep{0.1pt}
		\begin{tabular}{p{0.07\textwidth}p{0.065\textwidth}p{0.92\textwidth}}
            {\small Strokes} & {\small Input } & {\small Generated frames }   \\
			\includegraphics[width=0.0572\textwidth]{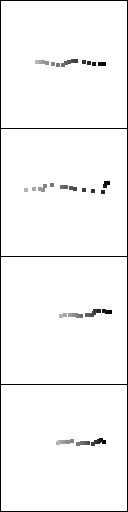}&
			\includegraphics[width=0.0572\textwidth]{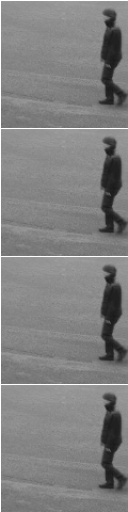}&
			\includegraphics[width=0.86\textwidth]{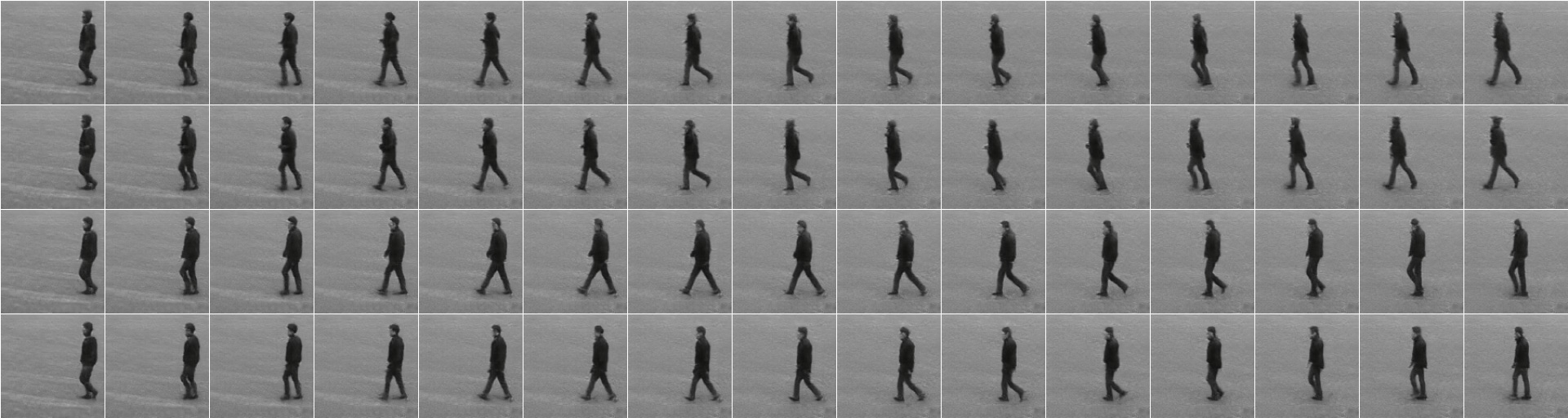}\\
		\end{tabular}
		\vspace{-1mm}
\end{subfigure}
    \caption{
    \label{fig:results-kth} 
    Experiments on the KTH dataset. 
    For each row, the first column is the input stroke where the intensity of the point goes from black to light grey over time. 
    The second column is the input image. 
    Starting from the third column to the end are the frames from the generated sequence.
    }
\end{figure*}

%% file: figures/long_sequences.tex
\begin{figure*}   
\setlength\tabcolsep{1pt}
\def\arraystretch{0.8}
\begin{tabular}{p{0.07\textwidth}p{0.07\textwidth}p{0.88\textwidth}}
{\small Strokes} & {\small Input } & {\small Generated frames }   \\
\includegraphics[width=0.065\textwidth]{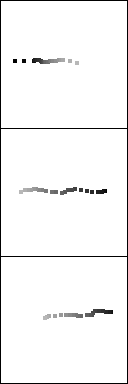}&
\includegraphics[width=0.065\textwidth]{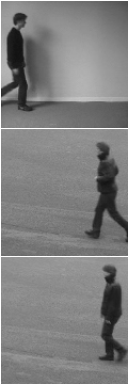}&
\includegraphics[width=0.85\textwidth]{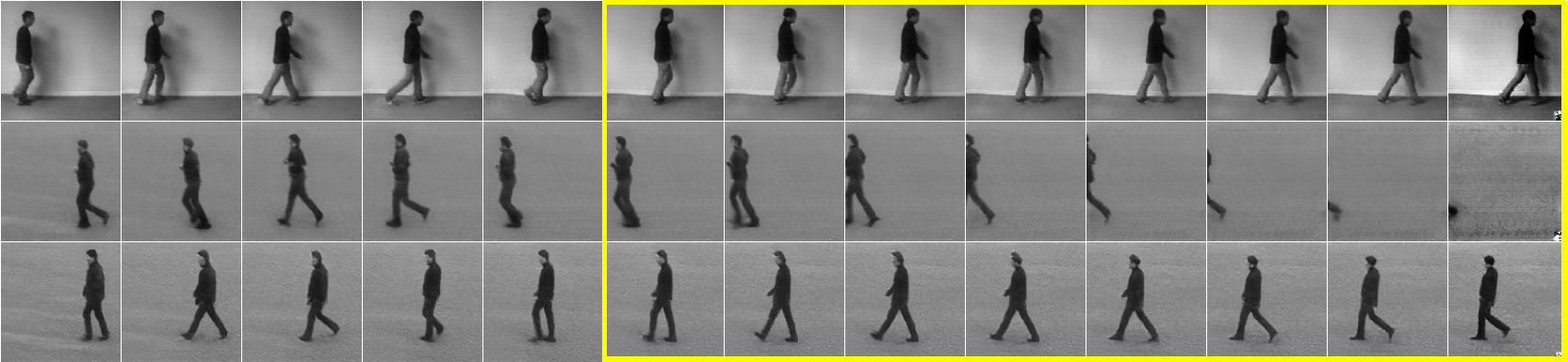}\\
\end{tabular}

\begin{tabular}{p{0.04\textwidth}p{0.05\textwidth}p{0.06\textwidth}p{0.05\textwidth}p{0.06\textwidth}p{0.06\textwidth}p{0.06\textwidth}p{0.06\textwidth}p{0.06\textwidth}p{0.06\textwidth}p{0.06\textwidth}p{0.06\textwidth}p{0.06\textwidth}p{0.06\textwidth}p{0.06\textwidth}p{0.06\textwidth}p{0.06\textwidth}p{0.06\textwidth}p{0.06\textwidth}}
&  & &  {\small t=3}&  \hspace{0.015\textwidth}{\small t=6}&   \hspace{0.02\textwidth}{\small t=9}& \hspace{0.02\textwidth}{\small t=12}&  \hspace{0.02\textwidth}{\small t=15}&  \hspace{0.02\textwidth}{\small t=16}& \hspace{0.02\textwidth}{\small t=17}&  \hspace{0.02\textwidth}{\small t=18}& \hspace{0.02\textwidth}{\small t=19}& \hspace{0.025\textwidth}{\small t=20}&  \hspace{0.022\textwidth}{\small t=21}& \hspace{0.022\textwidth} {\small t=22}& \hspace{0.023\textwidth}{\small t=23}\\
\end{tabular}
    \caption{Longer sequence. The frames inside the yellow boundary are predicted beyond the training regime with 16 frames. }
    \label{fig:results-kth-long-seq} 
\end{figure*}    

%% file: figures/comparison_li_endo.tex
\begin{figure*}  
	\newcommand{\width}{0.055\textwidth}
	\newcommand{\twidth}{0.04\textwidth}
	\centering
\begin{tabular}{cm{\width}m{\twidth}m{\twidth}m{\twidth}m{\twidth}m{\twidth}m{\twidth}m{\twidth}m{0\twidth}}
\multicolumn{1}{r}{Ground Truth}&
\includegraphics[width=\width]{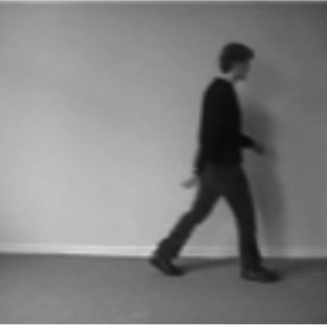}&
\includegraphics[width=\width]{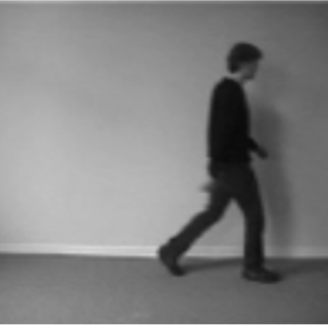}&
\includegraphics[width=\width]{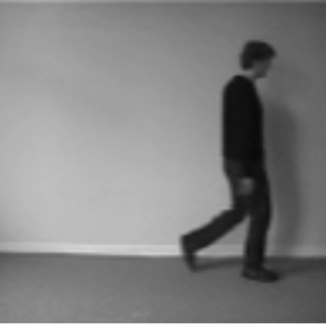}&
\includegraphics[width=\width]{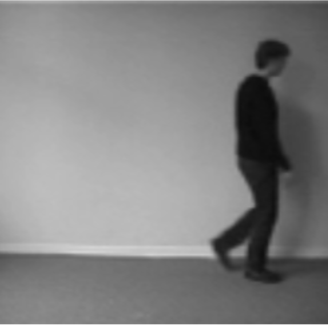}&
\includegraphics[width=\width]{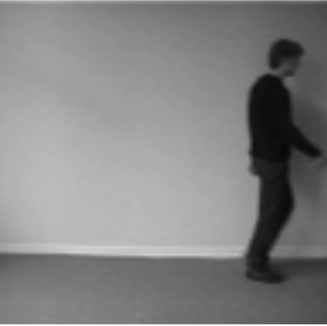}&
\includegraphics[width=\width]{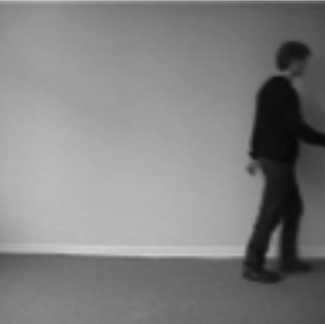}&
\includegraphics[width=\width]{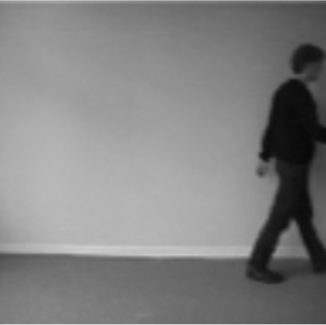}&
\includegraphics[width=\width]{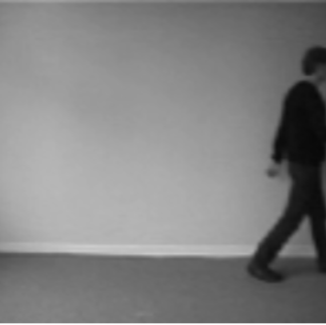}&
\includegraphics[width=\width]{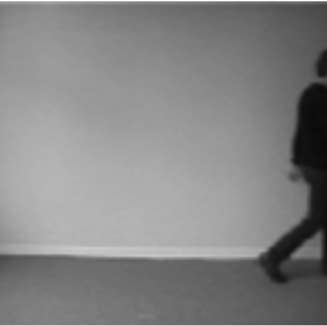}\\

\multicolumn{1}{r}{Li \etal~\cite{li2018flow}}& &
\includegraphics[width=\width]{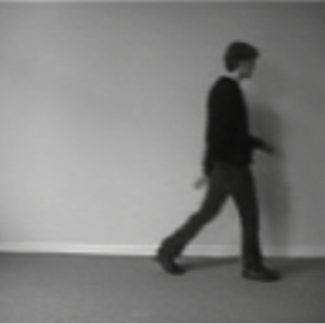}&
\includegraphics[width=\width]{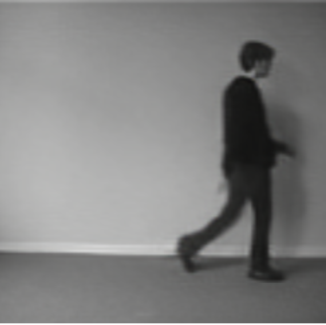}&
\includegraphics[width=\width]{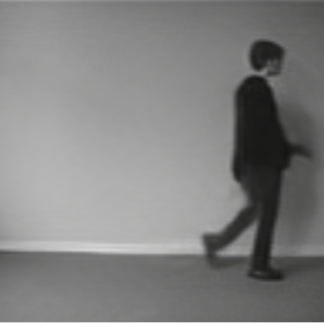}&
\includegraphics[width=\width]{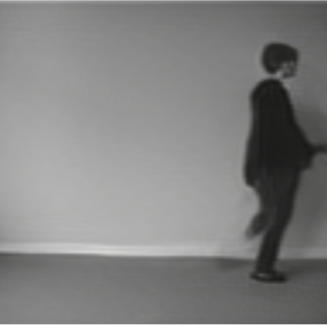}&
\includegraphics[width=\width]{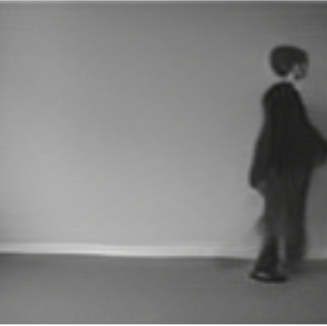}&
\includegraphics[width=\width]{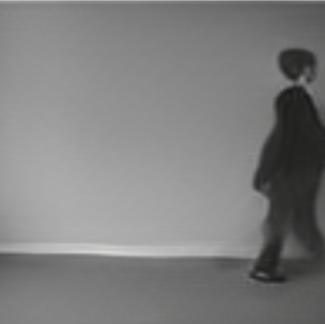}&
\includegraphics[width=\width]{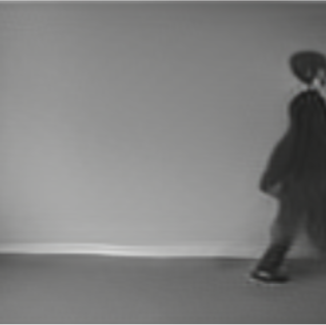}&
\includegraphics[width=\width]{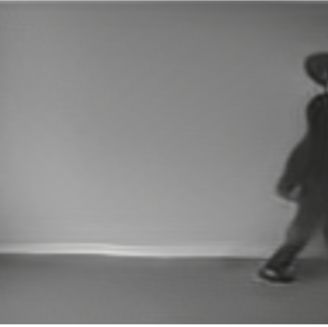}\\

\multicolumn{1}{r}{Endo \etal~\cite{Endo19}}& &
\includegraphics[width=\width]{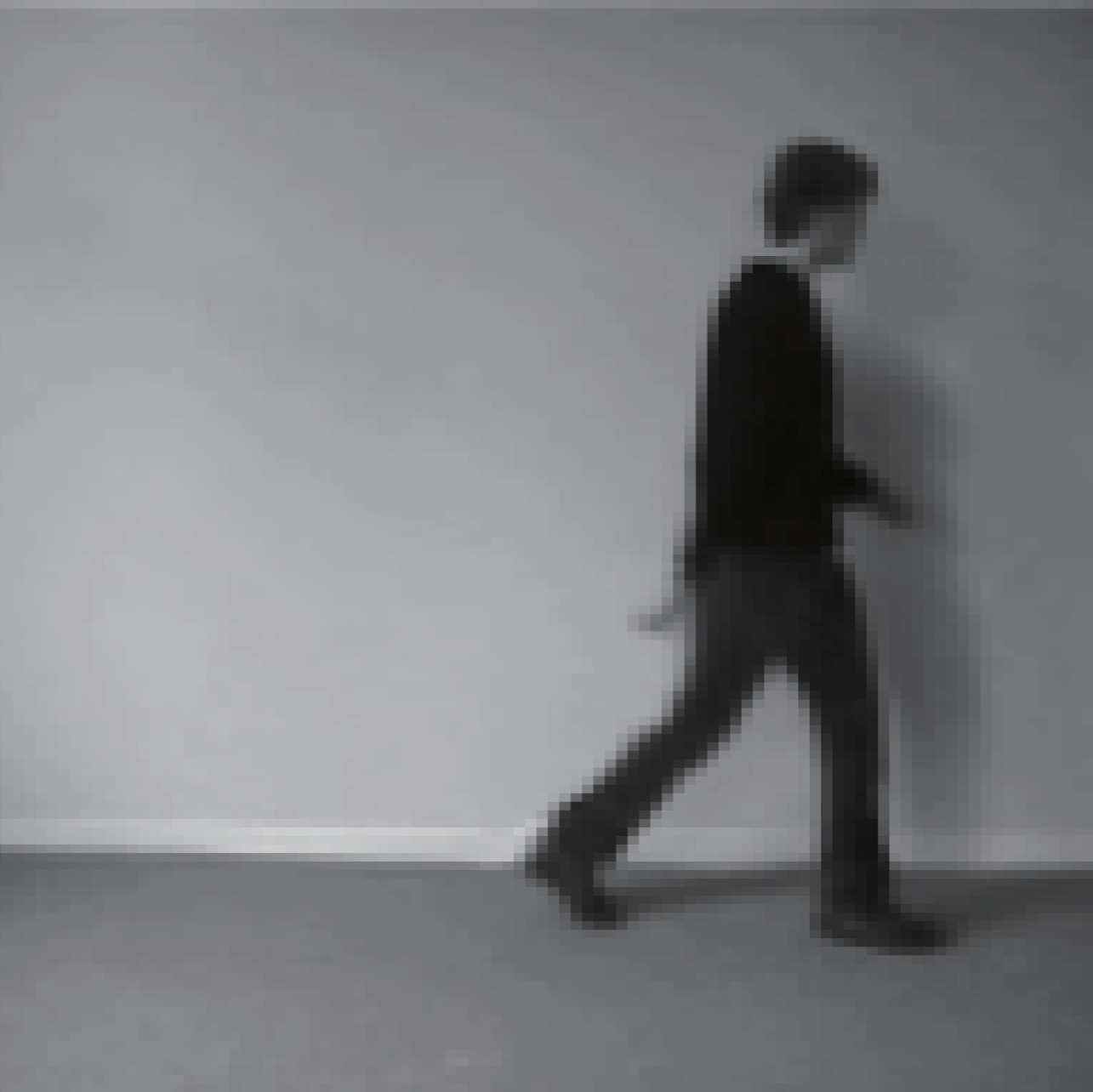}&
\includegraphics[width=\width]{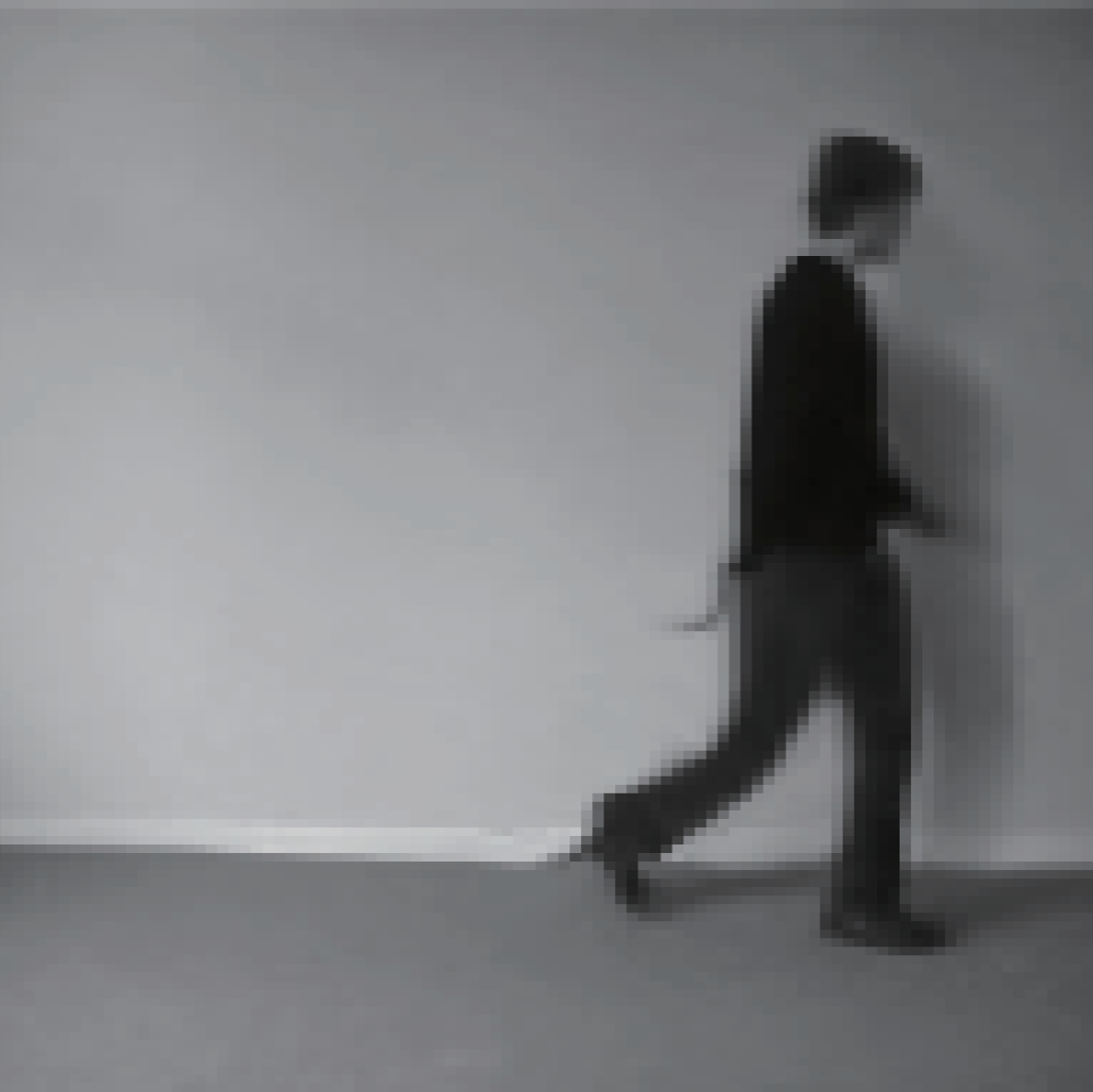}&
\includegraphics[width=\width]{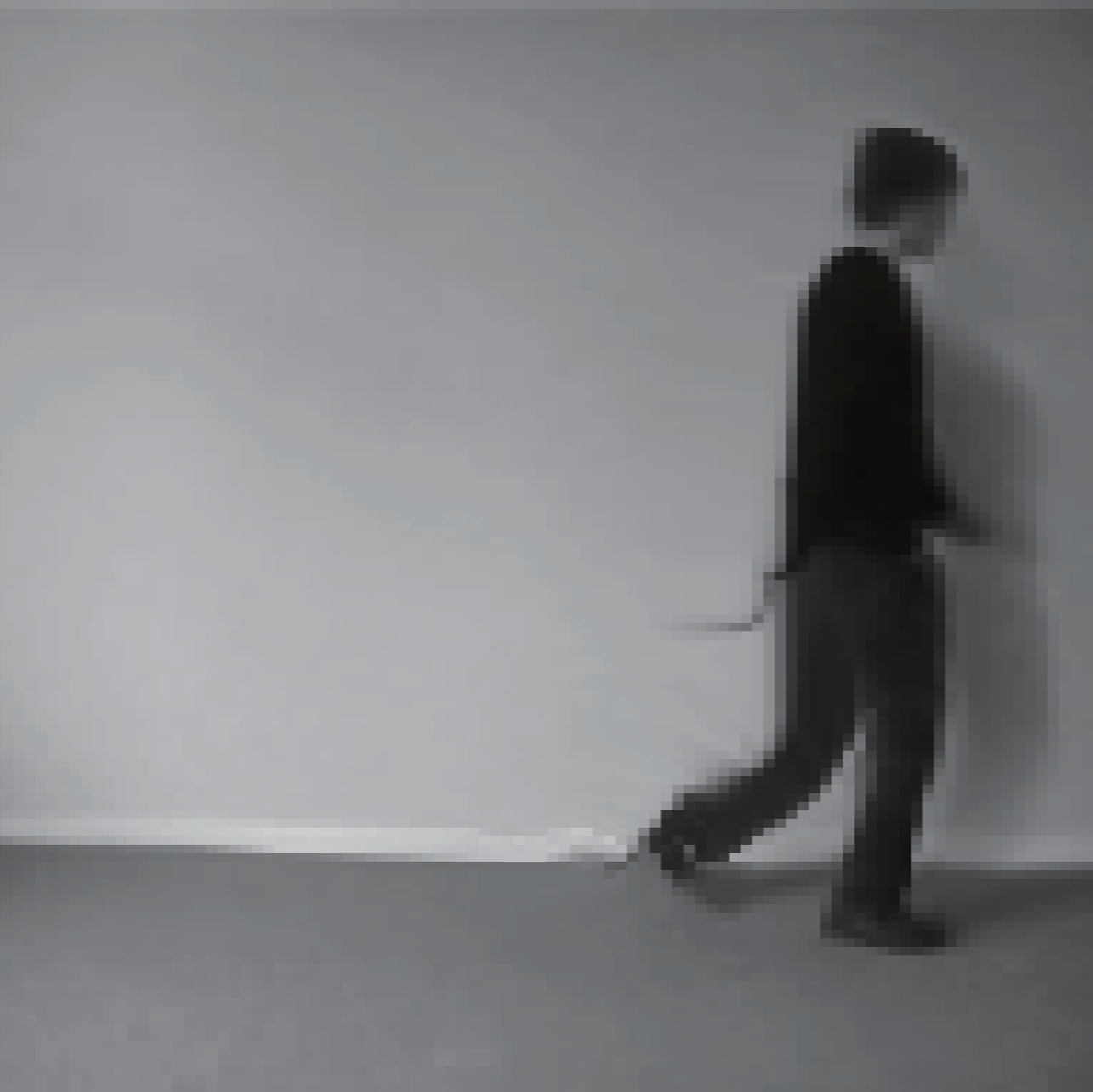}&
\includegraphics[width=\width]{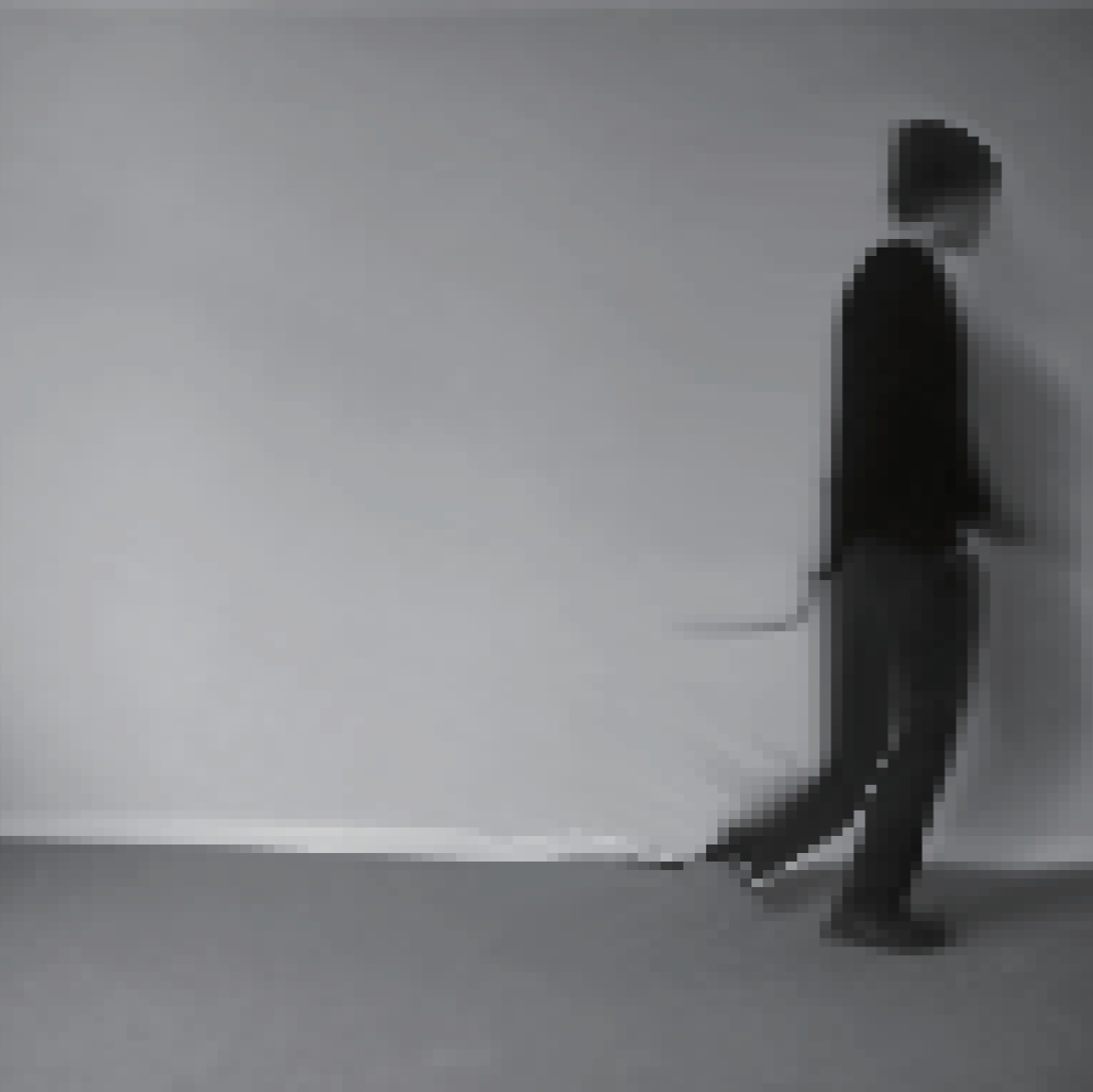}&
\includegraphics[width=\width]{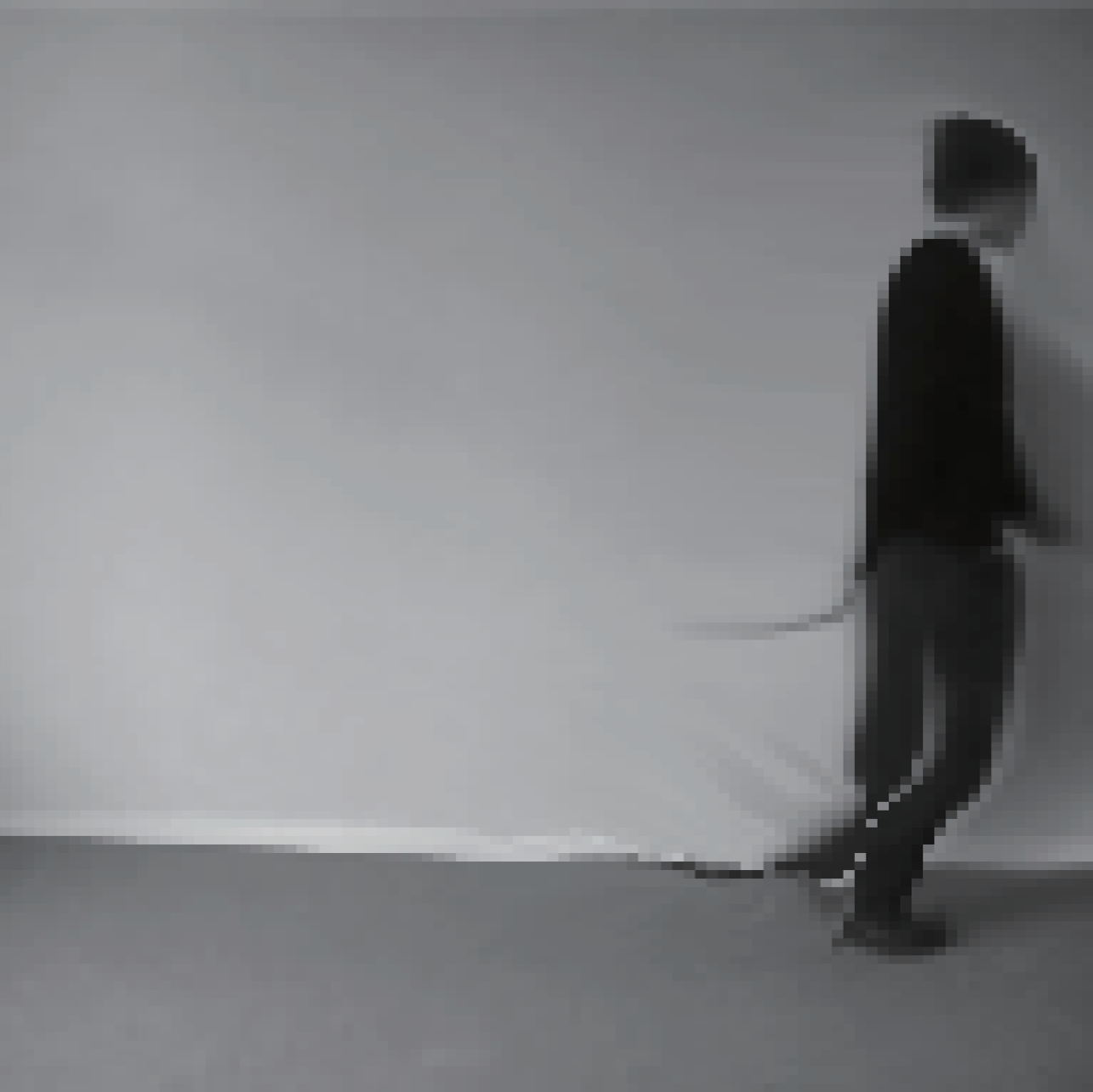}&
\includegraphics[width=\width]{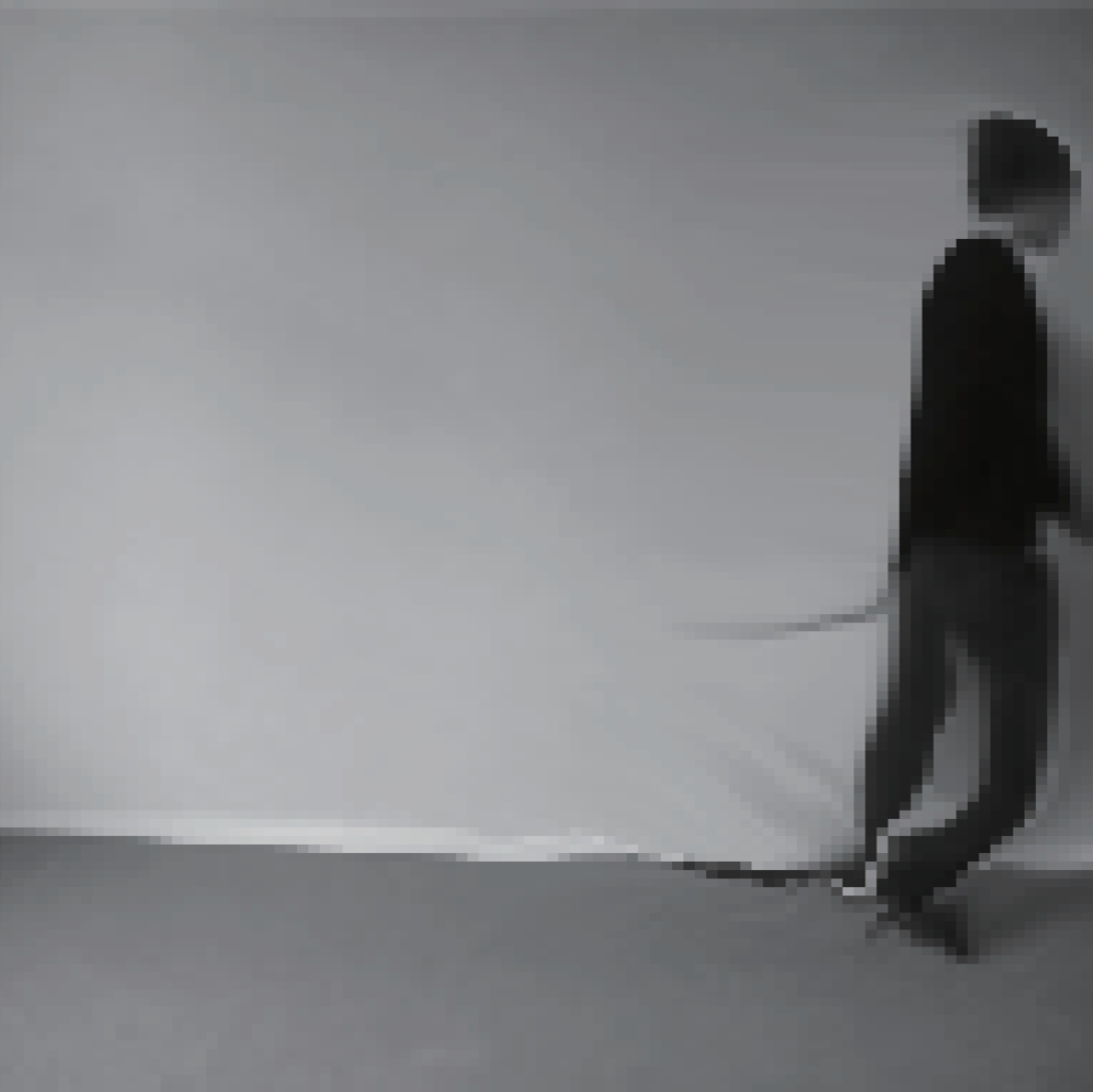}&
\includegraphics[width=\width]{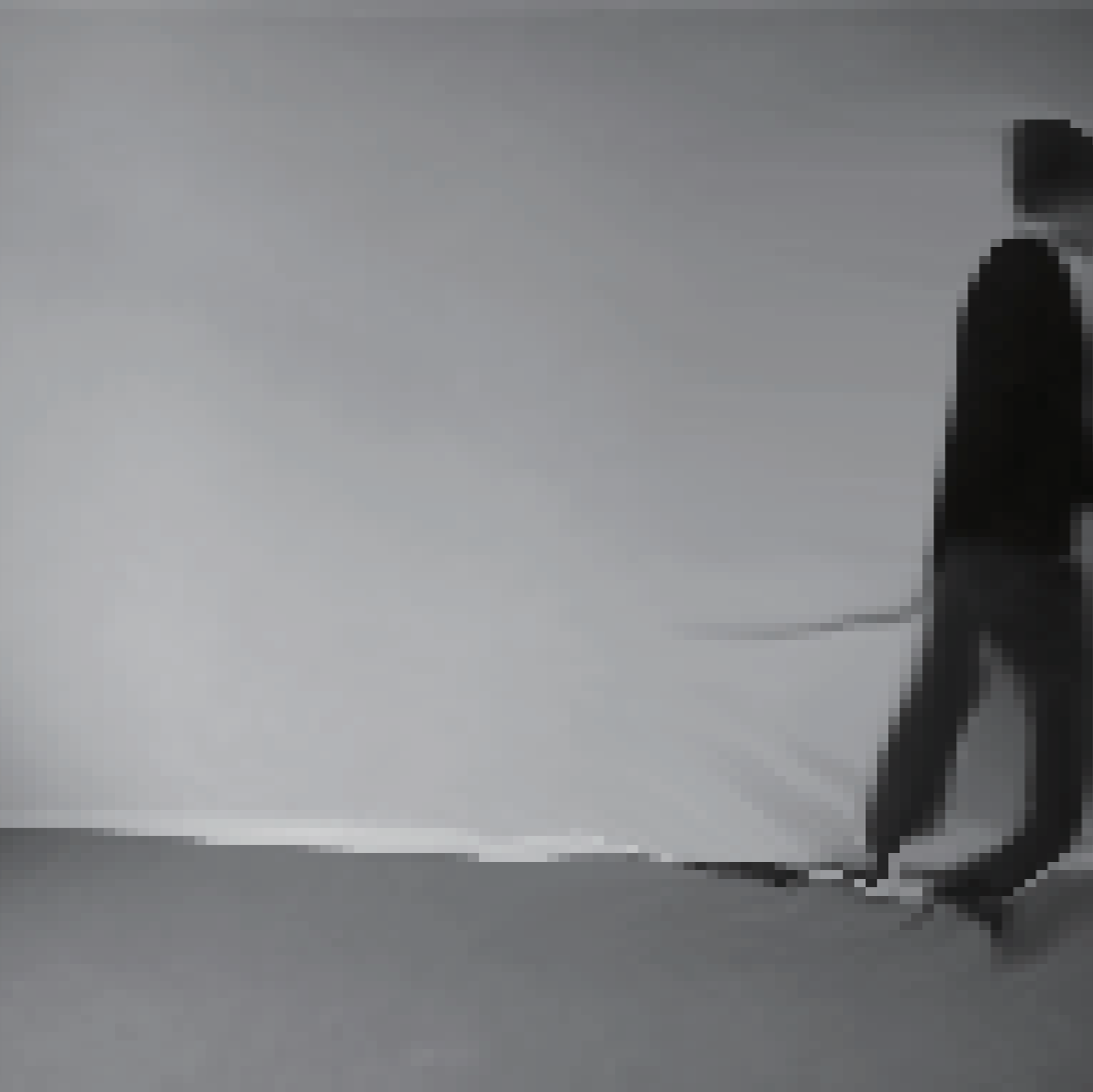}&
\includegraphics[width=\width]{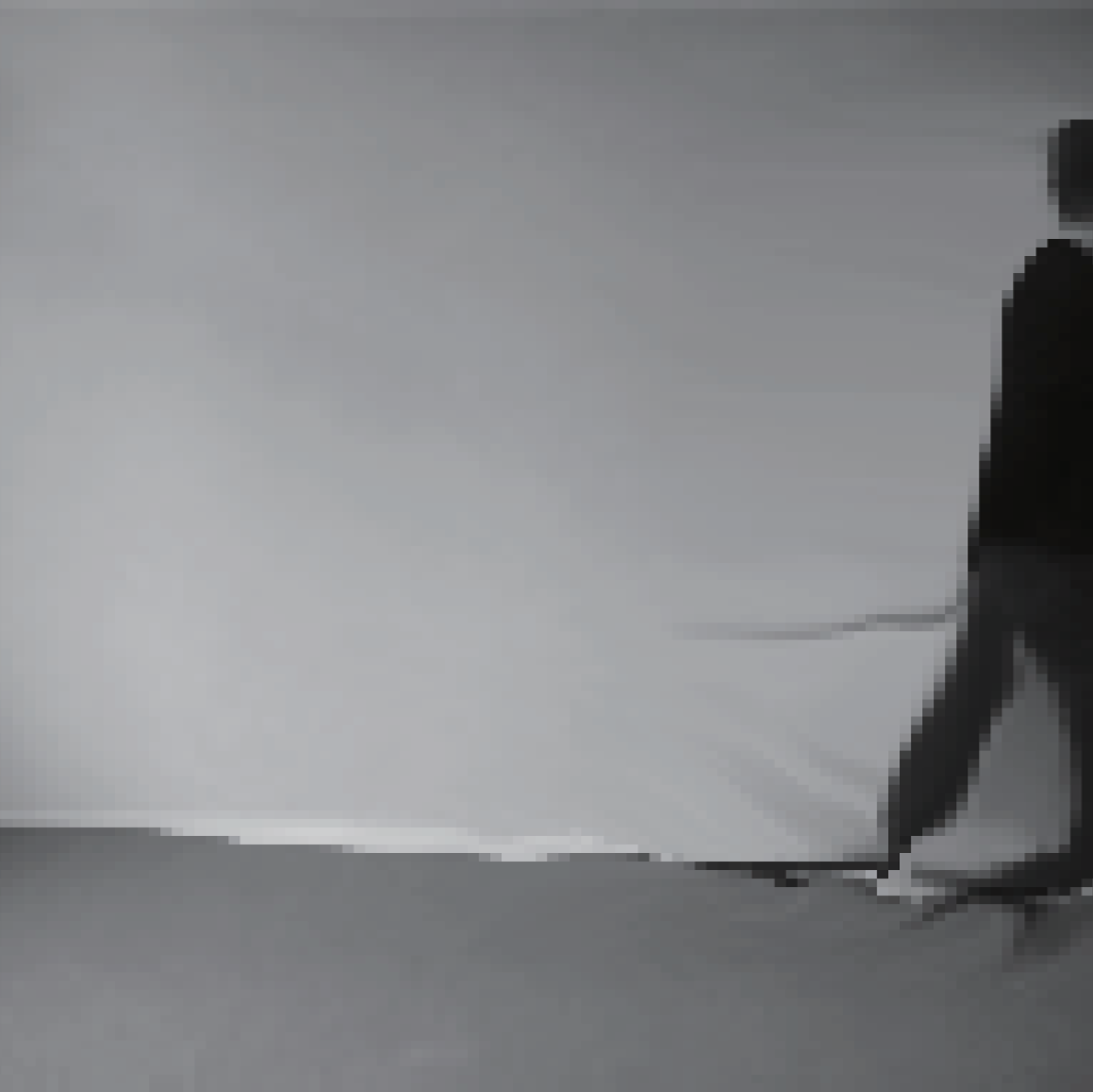}\\

\multicolumn{1}{r}{Ours}& &
\includegraphics[width=\width]{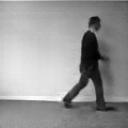}&
\includegraphics[width=\width]{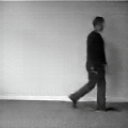}&
\includegraphics[width=\width]{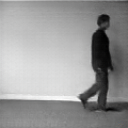}&
\includegraphics[width=\width]{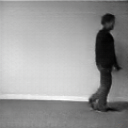}&
\includegraphics[width=\width]{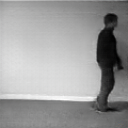}&
\includegraphics[width=\width]{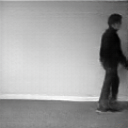}&
\includegraphics[width=\width]{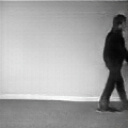}&
\includegraphics[width=\width]{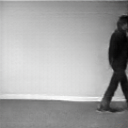}\\
 & \multicolumn{1}{c}{Input} 
 & \multicolumn{1}{c}{t=1} 
 & \multicolumn{1}{c}{t=3} 
 & \multicolumn{1}{c}{t=5} 
 & \multicolumn{1}{c}{t=7} 
 & \multicolumn{1}{c}{t=9} 
 & \multicolumn{1}{c}{t=11} 
 & \multicolumn{1}{c}{t=13} 
 & \multicolumn{1}{c}{t=15} 
\end{tabular}
    \caption{
     \label{fig:comparison-all}
     Qualitative comparison between Li \etal~\cite{li2018flow}, Endo \etal~\cite{Endo19}
     and ours on the KTH dataset. 
     The image size is $128 \times 128$ pixels. Our frames look sharp and more realistic than in the competing methods.
    }
\end{figure*}

%% file: figures/pose_smoothness.tex
\begin{table}
	\footnotesize 
	\caption{
		\label{tbl:quantitative-motion}
		Pose smoothness on generated KTH sequences. 
		We compare the rel.\@ mean Euclidean distance (\%) and std.\@ between pose joints in consecutive frames.
		The last column denotes the detection rate (\%) of valid bounding boxes across all classes. The percentage is relative to image size.
	}
	\centering
	\setlength{\tabcolsep}{4.5pt}
	\begin{tabular}{lrrrrrrr}  
		\toprule
		    											& \multicolumn{2}{c}{\emph{walking}} 	& \multicolumn{2}{c}{\emph{jogging}} 	& \multicolumn{2}{c}{\emph{running}} 	& 		det. \\
		\cmidrule(l){2-7}
		    											& mean 			& std 			& mean 	 		& std 			& mean 	 		& std 			&  rate\\
		\midrule
		Denton \etal~\cite{denton2018stochastic}		& 7.5			& 10.0			& 9.9	 		& 11.9			& 10.7 	 		& 11.5			& 54.2 \\
		Li \etal~\cite{li2018flow}  					& 7.4 			& 9.1			& 10.1 	 		& 11.3			& \textbf{ 8.7} & \textbf{9.9} 	& 54.9 \\	
		Ours											& \textbf{7.2}	& \textbf{7.7}	& \textbf{8.2}	& \textbf{9.1}	& 9.2  	 		& 10.5			& \textbf{87.1}\\
		Ground truth 									& 4.3 			& 5.7			& 5.3  	 		& 5.8			& 7.4  	 		& 6.8			& 100.0\\
		\bottomrule
	\end{tabular}
\end{table}

%% file: paper.bbl
\begin{thebibliography}{10}
\providecommand{\url}[1]{#1}
\csname url@samestyle\endcsname
\providecommand{\newblock}{\relax}
\providecommand{\bibinfo}[2]{#2}
\providecommand{\BIBentrySTDinterwordspacing}{\spaceskip=0pt\relax}
\providecommand{\BIBentryALTinterwordstretchfactor}{4}
\providecommand{\BIBentryALTinterwordspacing}{\spaceskip=\fontdimen2\font plus
\BIBentryALTinterwordstretchfactor\fontdimen3\font minus
  \fontdimen4\font\relax}
\providecommand{\BIBforeignlanguage}[2]{{%
\expandafter\ifx\csname l@#1\endcsname\relax
\typeout{** WARNING: IEEEtran.bst: No hyphenation pattern has been}%
\typeout{** loaded for the language `#1'. Using the pattern for}%
\typeout{** the default language instead.}%
\else
\language=\csname l@#1\endcsname
\fi
#2}}
\providecommand{\BIBdecl}{\relax}
\BIBdecl

\bibitem{schodl2000video}
A.~Sch{\"o}dl, R.~Szeliski, D.~H. Salesin, and I.~Essa, ``Video textures,'' in
  \emph{SIGGRAPH}, 2000, pp. 489--498.

\bibitem{yin2011shape}
C.~Yin, Y.~Gui, Z.~Xie, and L.~Ma, ``Shape context based video texture
  synthesis from still images,'' in \emph{ICCIS}.\hskip 1em plus 0.5em minus
  0.4em\relax IEEE, 2011, pp. 38--42.

\bibitem{goodfellow2014generative}
I.~Goodfellow, J.~Pouget-Abadie, M.~Mirza, B.~Xu, D.~Warde-Farley, S.~Ozair,
  A.~Courville, and Y.~Bengio, ``Generative adversarial nets,'' in \emph{NIPS},
  2014, pp. 2672--2680.

\bibitem{kingma2013auto}
D.~P. Kingma and M.~Welling, ``Auto-encoding variational bayes,''
  \emph{arXiv:1312.6114}, 2013.

\bibitem{saito2017temporal}
M.~Saito, E.~Matsumoto, and S.~Saito, ``Temporal generative adversarial nets
  with singular value clipping,'' in \emph{ICCV}, vol.~2, no.~3, 2017, p.~5.

\bibitem{vondrick2016generating}
C.~Vondrick, H.~Pirsiavash, and A.~Torralba, ``Generating videos with scene
  dynamics,'' in \emph{NIPS}, 2016, pp. 613--621.

\bibitem{mirza2014conditional}
M.~Mirza and S.~Osindero, ``Conditional generative adversarial nets,''
  \emph{arXiv:1411.1784}, 2014.

\bibitem{Pan_2019_CVPR}
J.~Pan, C.~Wang, X.~Jia, J.~Shao, L.~Sheng, J.~Yan, and X.~Wang, ``Video
  generation from single semantic label map,'' in \emph{CVPR}, June 2019.

\bibitem{van2016conditional}
A.~van~den Oord, N.~Kalchbrenner, L.~Espeholt, O.~Vinyals, A.~Graves
  \emph{et~al.}, ``Conditional image generation with pixelcnn decoders,'' in
  \emph{NIPS}, 2016, pp. 4790--4798.

\bibitem{hochreiter1997long}
S.~Hochreiter and J.~Schmidhuber, ``Long short-term memory,'' \emph{Neural
  computation}, vol.~9, no.~8, pp. 1735--1780, 1997.

\bibitem{tulyakov2018mocogan}
S.~Tulyakov, M.-Y. Liu, X.~Yang, and J.~Kautz, ``{MoCoGAN}: Decomposing motion
  and content for video generation,'' in \emph{CVPR}, 2018, pp. 1526--1535.

\bibitem{byeon2018contextvp}
W.~Byeon, Q.~Wang, R.~Kumar~Srivastava, and P.~Koumoutsakos, ``Contextvp: Fully
  context-aware video prediction,'' in \emph{ECCV}, 2018, pp. 753--769.

\bibitem{pascanu2013difficulty}
R.~Pascanu, T.~Mikolov, and Y.~Bengio, ``On the difficulty of training
  recurrent neural networks,'' in \emph{ICML}, 2013, pp. 1310--1318.

\bibitem{mathieu2015deep}
M.~Mathieu, C.~Couprie, and Y.~LeCun, ``Deep multi-scale video prediction
  beyond mean square error,'' \emph{arXiv:1511.05440}, 2015.

\bibitem{denton2018stochastic}
E.~Denton and R.~Fergus, ``Stochastic video generation with a learned prior,''
  in \emph{ICML}, vol.~80, 10--15 Jul 2018, pp. 1174--1183.

\bibitem{denton2017_disentagled}
E.~L. Denton and v.~Birodkar, ``Unsupervised learning of disentangled
  representations from video,'' in \emph{NIPS}, 2017, pp. 4414--4423.

\bibitem{Wang_2019_ICCV}
T.-H. Wang, Y.-C. Cheng, C.~H. Lin, H.-T. Chen, and M.~Sun, ``Point-to-point
  video generation,'' in \emph{ICCV}, October 2019.

\bibitem{babaeizadeh2018stochastic}
M.~Babaeizadeh, C.~Finn, D.~Erhan, R.~H. Campbell, and S.~Levine, ``Stochastic
  variational video prediction,'' in \emph{ICLR}, 2018.

\bibitem{hao2018controllable}
Z.~Hao, X.~Huang, and S.~Belongie, ``Controllable video generation with sparse
  trajectories,'' in \emph{CVPR}, 2018, pp. 7854--7863.

\bibitem{he2018probabilistic}
J.~He, A.~Lehrmann, J.~Marino, G.~Mori, and L.~Sigal, ``Probabilistic video
  generation using holistic attribute control,'' in \emph{ECCV}, 2018, pp.
  452--467.

\bibitem{li2018flow}
Y.~Li, C.~Fang, J.~Yang, Z.~Wang, X.~Lu, and M.-H. Yang, ``Flow-grounded
  spatial-temporal video prediction from still images,'' in \emph{ECCV}, 2018.

\bibitem{xiong2018learning}
W.~Xiong, W.~Luo, L.~Ma, W.~Liu, and J.~Luo, ``Learning to generate time-lapse
  videos using multi-stage dynamic generative adversarial networks,'' in
  \emph{CVPR}, 2018, pp. 2364--2373.

\bibitem{Endo19}
Y.~Endo, Y.~Kanamori, and S.~Kuriyama, ``Animating landscape: Self-supervised
  learning of decoupled motion and appearance for single-image video
  synthesis,'' \emph{ACM Trans. Graph.}, vol.~38, no.~6, pp. 175:1--175:19,
  Nov. 2019.

\bibitem{cao2017realtime}
Z.~Cao, T.~Simon, S.-E. Wei, and Y.~Sheikh, ``Realtime multi-person 2d pose
  estimation using part affinity fields,'' in \emph{CVPR}, 2017, pp.
  7291--7299.

\bibitem{newell2016stacked}
A.~Newell, K.~Yang, and J.~Deng, ``Stacked hourglass networks for human pose
  estimation,'' in \emph{ECCV}.\hskip 1em plus 0.5em minus 0.4em\relax
  Springer, 2016, pp. 483--499.

\bibitem{wei2016convolutional}
S.-E. Wei, V.~Ramakrishna, T.~Kanade, and Y.~Sheikh, ``Convolutional pose
  machines,'' in \emph{CVPR}, 2016, pp. 4724--4732.

\bibitem{balakrishnan2018synthesizing}
G.~Balakrishnan, A.~Zhao, A.~V. Dalca, F.~Durand, and J.~Guttag, ``Synthesizing
  images of humans in unseen poses,'' in \emph{CVPR}, 2018, pp. 8340--8348.

\bibitem{chan2019everybody}
C.~Chan, S.~Ginosar, T.~Zhou, and A.~A. Efros, ``Everybody dance now,'' in
  \emph{ICCV}, 2019, pp. 5933--5942.

\bibitem{davis2003sketching}
J.~Davis, M.~Agrawala, E.~Chuang, Z.~Popovi{\'c}, and D.~Salesin, ``A sketching
  interface for articulated figure animation,'' in \emph{SIGGRAPH}, 2003, pp.
  320--328.

\bibitem{thorne2004motion}
M.~Thorne, D.~Burke, and M.~van~de Panne, ``Motion doodles: an interface for
  sketching character motion,'' in \emph{ACM Trans. Graph.}, vol.~23, no.~3,
  2004, pp. 424--431.

\bibitem{chen2005character}
B.-Y. Chen, Y.~Ono, and T.~Nishita, ``Character animation creation using
  hand-drawn sketches,'' \emph{The Visual Computer}, vol.~21, no. 8-10, pp.
  551--558, 2005.

\bibitem{johnson2016perceptual}
J.~Johnson, A.~Alahi, and L.~Fei-Fei, ``Perceptual losses for real-time style
  transfer and super-resolution,'' in \emph{ECCV}, 2016, pp. 694--711.

\bibitem{simonyan2014very}
K.~Simonyan and A.~Zisserman, ``Very deep convolutional networks for
  large-scale image recognition,'' \emph{ICLR}, 2015.

\bibitem{huang2017densely}
G.~Huang, Z.~Liu, L.~Van Der~Maaten, and K.~Q. Weinberger, ``Densely connected
  convolutional networks.'' in \emph{CVPR}, vol.~1, no.~2, 2017, p.~3.

\bibitem{miyato2018spectral}
T.~Miyato, T.~Kataoka, M.~Koyama, and Y.~Yoshida, ``Spectral normalization for
  generative adversarial networks,'' \emph{ICLR}, 2018.

\bibitem{kingma2014adam}
D.~P. Kingma and J.~Ba, ``Adam: A method for stochastic optimization,''
  \emph{arXiv:1412.6980}, 2014.

\bibitem{lecun1998gradient}
Y.~LeCun, L.~Bottou, Y.~Bengio, and P.~Haffner, ``Gradient-based learning
  applied to document recognition,'' \emph{Proceedings of the IEEE}, vol.~86,
  no.~11, pp. 2278--2324, 1998.

\bibitem{pushdataset}
C.~Finn, I.~Goodfellow, and S.~Levine, ``Unsupervised learning for physical
  interaction through video prediction,'' \emph{NIPS}, 2016.

\bibitem{schuldt2004recognizing}
C.~Schuldt, I.~Laptev, and B.~Caputo, ``Recognizing human actions: a local svm
  approach,'' in \emph{ICPR}, vol.~3.\hskip 1em plus 0.5em minus 0.4em\relax
  IEEE, 2004, pp. 32--36.

\bibitem{ActionsAsSpaceTimeShapes_pami07}
L.~Gorelick, M.~Blank, E.~Shechtman, M.~Irani, and R.~Basri, ``Actions as
  space-time shapes,'' \emph{TPAMI}, vol.~29, no.~12, pp. 2247--2253, December
  2007.

\bibitem{ionescu2014human3}
C.~Ionescu, D.~Papava, V.~Olaru, and C.~Sminchisescu, ``Human3. 6m: Large scale
  datasets and predictive methods for 3d human sensing in natural
  environments,'' \emph{TPAMI}, vol.~36, no.~7, pp. 1325--1339, 2014.

\bibitem{redmon2016you}
J.~Redmon, S.~Divvala, R.~Girshick, and A.~Farhadi, ``You only look once:
  Unified, real-time object detection,'' in \emph{CVPR}, 2016, pp. 779--788.

\bibitem{ssim}
Z.~Wang, A.~C. Bovik, H.~R. Sheikh, and E.~P. Simoncelli, ``Image quality
  assessment: From error visibility to structural similarity,'' \emph{Trans.
  Img. Proc.}, vol.~13, no.~4, pp. 600--612, 2004.

\bibitem{zhang2018unreasonable}
R.~Zhang, P.~Isola, A.~A. Efros, E.~Shechtman, and O.~Wang, ``The unreasonable
  effectiveness of deep features as a perceptual metric,'' in \emph{CVPR},
  2018.

\end{thebibliography}
